\documentclass[12pt,preprint]{aastex}

\shorttitle{A Method of Constructing Riemann Ellipsoids}

\shortauthors{ Ou }

\begin{document}

%% LaTeX will automatically break titles if they run longer than
%% one line. However, you may use \\ to force a line break if
%% you desire.

%%\title{{A Method for Constructing Riemann and Riemann-like Ellipsoidal Configurations}}
\title{{An Approximate Solver for Riemann and Riemann-like Ellipsoidal Configurations}}

%% Use \author, \affil, and the \and command to format
%% author and affiliation information.
%% Note that \email has replaced the old \authoremail command
%% from AASTeX v4.0. You can use \email to mark an email address
%% anywhere in the paper, not just in the front matter.
%% As in the title, use \\ to force line breaks.

\author{Shangli Ou}
\affil{Department of Physics \& Astronomy and Center for Computation \& Technology, \\
      Louisiana State University, Baton Rouge, LA  70803}

%% Mark off your abstract in the ``abstract'' environment. In the manuscript
%% style, abstract will output a Received/Accepted line after the
%% title and affiliation information. No date will appear since the author
%% does not have this information. The dates will be filled in by the
%% editorial office after submission.

\begin{abstract}

We introduce a new technique for constructing three-dimensional (3D) models
of incompressible Riemann S-type ellipsoids and compressible triaxial
configurations that share the same velocity field as that of
 Riemann S-type ellipsoids. 
Our incompressible models are exact steady-state configurations; 
our compressible models represent approximate steady-state 
equilibrium configurations.
Models built from this method can be used to study a variety of relevant 
astrophysical and geophysical problems. 

\end{abstract}

%% Keywords should appear after the \end{abstract} command. The uncommented
%% example has been keyed in ApJ style. See the instructions to authors
%% for the journal to which you are submitting your paper to determine
%% what keyword punctuation is appropriate.

%% Authors who wish to have the most important objects in their paper
%% linked in the electronic edition to a data center may do so in the
%% subject header.  Objects should be in the appropriate "individual"
%% headers (e.g. quasars: individual, stars: individual, etc.) with the
%% additional provision that the total number of headers, including each
%% individual object, not exceed six.  The \objectname{} macro, and its
%% alias \object{}, is used to mark each object.  The macro takes the object
%% name as its primary argument.  This name will appear in the paper
%% and serve as the link's anchor in the electronic edition if the name
%% is recognized by the data centers.  The macro also takes an optional
%% argument in parentheses in cases where the data center identification
%% differs from what is to be printed in the paper.


%% From the front matter, we move on to the body of the paper.
%% In the first two sections, notice the use of the natbib \citep
%% and \citet commands to identify citations.  The citations are
%% tied to the reference list via symbolic KEYs. The KEY corresponds
%% to the KEY in the \bibitem in the reference list below. We have
%% chosen the first three characters of the first author's name plus
%% the last two numeral of the year of publication as our KEY for
%% each reference.

\section{Introduction}

Our understanding of stellar structure and stability has been built 
upon classical 
analytical studies of equilibrium configurations of incompressible
self-gravitating systems by Maclaurin, Jacobi, Dedekind, Darwin, Riemann,
and especially \cite{Ch69}.
With the help of modern computers and numerical methods,
such as the self-consistent-field (SCF) technique \citep{OM68, H86A}
and hydrodynamics techniques,
these studies have been extended to compressible configurations
by a variety of researchers
\citep{TDM85, H86A, H86B, WT88,CT00, SKE02}.
However, most three-dimensional (3D) hydrodynamical studies 
have been confined to a limited parameter 
space, i.e., starting from two-dimensional (2D) axisymmetric equilibrium models
(compressible analogues of Maclaurin spheroids)
because we have not been able to build 3D compressible equilibrium models with
complicated flows.  
To the author's knowledge, 
techniques have only been developed to build compressible equilibrium models 
of nonaxisymmetric configurations for a few systems with simplified rotational profiles,
e.g., rigidly rotating systems \citep{HE84,H86B}, irrotational systems \citep{UE98},
and configurations that are stationary in the inertial frame \citep{UE96}.
Our aim is to find a method by which a much wider class of nonaxisymmetric
configurations can be readily constructed. This will permit us to more
clearly understand the viability of evolutionary sequences 
of triaxial configurations and will provide models 
whose stability properties can be readily analyzed.

Previous linear stability analyses \citep{Ch69,LL93,LL96,LS99} 
have revealed a number of hydrodynamical instabilities 
that might arise in incompressible Riemann ellipsoids, 
but no corresponding numerical work 
using state-of-the-art 3D hydrodynamic techniques has been
carried out.
%%due to our inability to construct 3D models for these configurations.
In particular, Riemann S-type ellipsoids have been found
to be subject to a hydrodynamic strain instability associated with
elliptical stream lines \citep{LL96},
which raises concerns about the stability of certain types of
geophysical flows and leads to suspicions about the evolutionary path of 
stars that are driven by gravitational-radiation-reaction (GRR) 
forces toward the Dedekind-sequence.
In a recent nonlinear study of the secular bar-mode instability
induced by GRR \citep{OTL04},
a uniformly rotating, axisymmetric neutron star,
which was secularly unstable and driven by artifically enhanced GRR, 
evolved into a bar-like configuration with a very slow pattern speed.
However, this special bar configuration became
unstable to some kind of turbulence-like instability 
while evolving toward the Dedekind-sequence,
that is, toward a stationary ellipsoidal structure in the inertial frame
maintained purely by the internal motion of the fluid.
It is suspected that this turbulence-like instability may
be the elliptical strain instability identified in the
earlier linear stability analyses. 
But no definite conclusion can be drawn 
because the stability of only one Dedekind-like configuration was studied. 
Following \cite{OTL04}, one might consider generating additional 
ellipsoidal structures hydrodynamically,  
but the generation of each additional ellipsoidal model would be expensive
in the sense that one has to evolve an initially axisymmetric model 
for a very long time, even with
artificially enhanced GRR \citep{SK04,OTL04}.
This approach would make it impractical to undertake a full stability
investigation through the entire parameter space of ellipsoidal models.

Here we present a new 3D SCF technique that is capable of building the full range
of incompressible Riemann S-type ellipsoids with nontrivial internal flows 
(i.e., not just the uniformly rotating Jacobi configurations, 
 which are thought to be the end point of secularly unstable stars driven
 by viscosity, and stationary Dedekind configurations)
as well as compressible triaxial models
that share the same divergence-free velocity field as Riemann S-type ellipsoids. 
Our compressible models satisfy the steady-state Euler's equation exactly,
but only satisfy the steady-state continuity equation approximately,
hence they are only in quasi-equilibrium. 
However, the violation of the steady-state continuity equation is very small
for nearly incompressible models, 
hence these models can be used effectively as initial configurations for
a hydrodynamical study of the stability of compressible analogues of 
Riemann S-type ellipsoids.
We note that, to date, hydrodynamical techniques have not been used 
to examine the stability of Riemann ellipsoids 
because the sound speed is formally infinite for incompressible models,
while the time step in hydrodynamic codes is usually inversely 
proportional to the sound speed.
On the other hand, 
since the elliptical strain instability associated with elliptical stream lines
would set in on a dynamical time scale,
we believe that our new ``quasi-equilibrium'' models will prove 
to be good starting points for hydrodynamical studies of this instability.

Our numerical method inherits one central idea of the traditional SCF method in that
it makes use of the properties of the velocity field
of Riemann S-type ellipsoids to turn the vector form of the steady-state
Euler's equation into a scalar form, thus yielding a uniform Bernoulli's 
function. This enables us to follow the procedures of the traditional SCF 
method to use
several boundary conditions to determine quantities elsewhere.
To be concise, we will refer readers to \cite{H86A} and references therein
 for details of the traditional
SCF technique, and focus on the new aspects of our new method.

\section{Numerical Methods}

\begin{figure}[ht]
\epsscale{1.0} \plotone{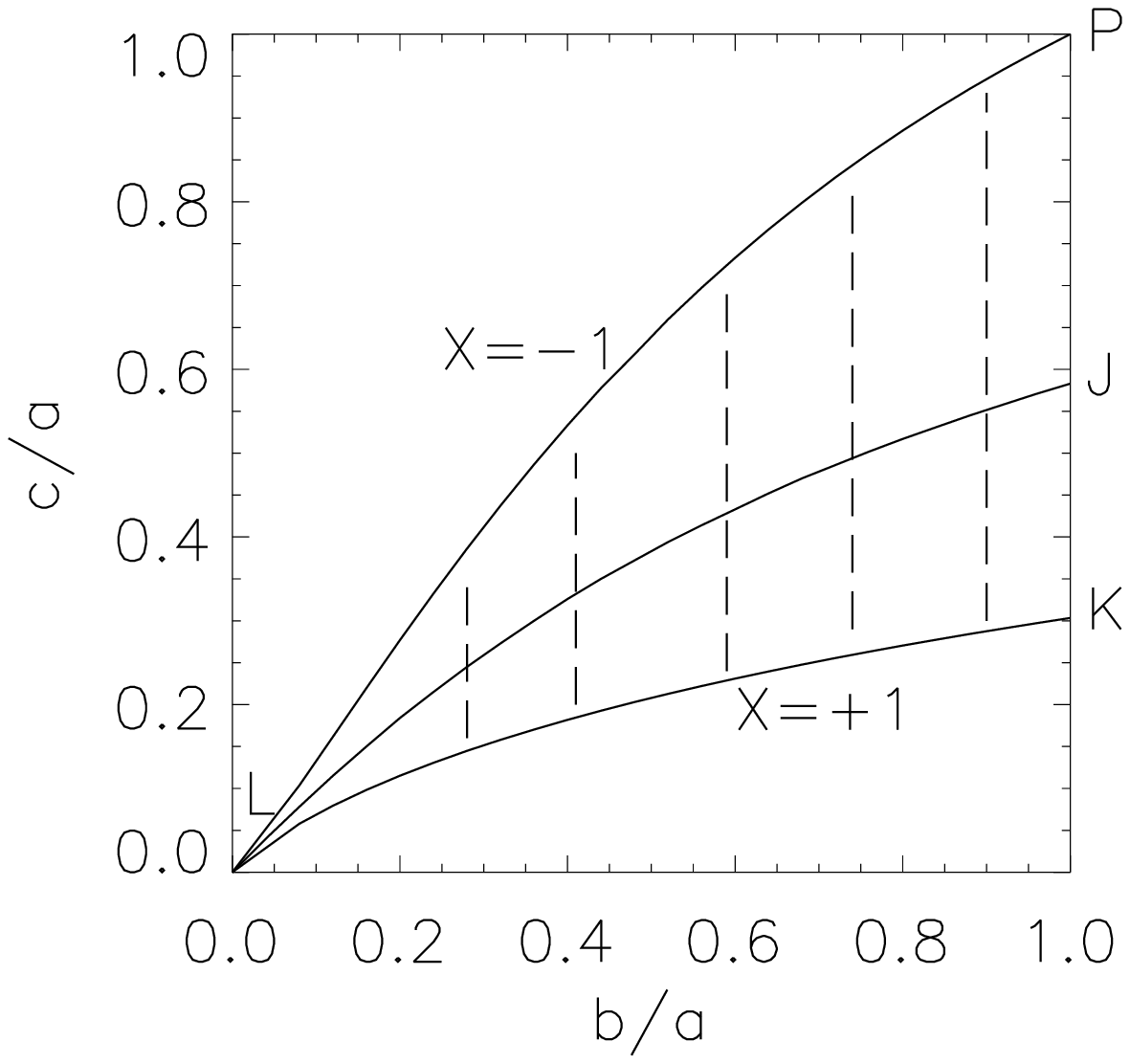}
\caption{ Parameter space of Riemann S-type ellipsoids 
     on a ($b/a$, $c/a$) plot. 
     Curve LJ denotes for the Jacobi/Dedekind sequence,
     curves LP and LK are the two self-adjoint sequences with $X=\mp 1$.
     Models are constructed along dashed lines in order to cover most of the
     parameter space.
     \label{triaxial}}
\end{figure}

Figure \ref{triaxial} reproduces key elements of a plot 
that \citet[see his chapter 7, figure 15]{Ch69}
used to discuss sequences of Riemann S-type ellipsoids.
In this ($b/a$,$c/a$) parameter plot, 
$a$, $b$, and $c$ are the three principal semiaxes of an ellipsoid,
and each Riemann S-type ellipsoid with angular velocity $\omega$ and 
uniform vorticity $\zeta$ %%(only has z component) 
in the rotating frame resides between 
the two self-adjoint sequences (curves LP and LK) with $X=\mp 1$, 
where
\begin{eqnarray}
   X &\equiv& f \frac{ab}{a^2+b^2} \,, \label{eqn:X}\\
   f&\equiv&\frac{\zeta}{\omega} \,. \label{eqn:f}
\end{eqnarray}
The velocity field of a Riemann S-type ellipsoid as viewed 
from a frame rotating with angular velocity $\omega$
 takes the following form:  
\begin{equation}
  \vec{v} = \lambda (ay/b, -bx/a, 0) \,, \label{vel}
\end{equation}
where $\lambda$ is a constant that determines the magnitude of the
internal motion of the fluid, and the origin of the x-y coordinate
system is at the center of the ellipsoid.
This velocity field $\vec{v}$ is designed so that velocity vectors everywhere
are always aligned with elliptical stream lines by demanding that they be
tangent to the equipotential contours, which are
concentric ellipses. %%with the same axis ratio. 

Now, let's turn to a general self-gravitating system with some compressibility. 
If we adopt a polytropic equation of state, $p=k\rho^{1+1/n}$
($p$ is pressure, $k$ is the polytropic constant and
 $n$ is the polytropic index),
Euler's equation for steady-state flows in the rotating frame is,
\begin{equation}
  \vec{v} \cdot \nabla(\vec{v})=
        -\nabla[ H+\Phi-\frac{1}{2}\omega^2(x^2+y^2) ]
        - 2 \vec{\omega} \times \vec{v} \,,  \label{euler}
\end{equation}
%%where $\vec{v}$ is the three-velocity of the fluid, 
where H is enthalpy, and
$\Phi$ is the gravitational potential.
With the velocity field specified by Eq. (\ref{vel}), one can show that 
Eq. (\ref{euler}) is equivalent to the following equation, 
\begin{eqnarray}
  -\nabla[\frac{1}{2} \lambda^2 (x^2+y^2)] =
     -\nabla[ H+\Phi-\frac{1}{2}\omega^2 (x^2+y^2) ] \nonumber\\
     -\nabla[ \omega \lambda(\frac{b}{a}x^2 + \frac{a}{b}y^2) ] \,.
\end{eqnarray}
Hence, within the configuration the following Bernoulli's function
must be uniform in space:
\begin{eqnarray}
  H+\Phi-\frac{1}{2} \omega^2(x^2+y^2)-\frac{1}{2}\lambda^2(x^2+y^2) \nonumber \\
   + \omega \lambda (\frac{b}{a}x^2 + \frac{a}{b}y^2) = C_1  \label{bernoulli} \,,
\end{eqnarray}
where $C_1$ is a constant.
It should be mentioned that in this Bernoulli's function,
$\omega$ and $\lambda$ are interchangeable. This reflects the fact
that, for any direct configuration ($\omega>\lambda$) 
in which fluid motion is dominated by figure rotation, 
the adjoint configuration ($\omega<\lambda$) 
in which fluid motion is dominated by internal motions can be
obtained by simply interchanging $\omega$ and $\lambda$. 
We find it is useful to define an effective potential $\Phi_{\mathrm{eff}}$ as:
\begin{equation}
  \Phi_{\mathrm{eff}} \equiv \Phi-\frac{1}{2} \omega^2(x^2+y^2)-\frac{1}{2}\lambda^2(x^2+y^2) \nonumber \\
   + \omega \lambda (\frac{b}{a}x^2 + \frac{a}{b}y^2) \label{phieff} \,,
\end{equation}
hence, the Bernoulli's function can also be written as:
\begin{equation}
  H+\Phi_{\mathrm{eff}} = C_1 \,.
\end{equation}
In our later discussion of equi-potential (iso-density) surfaces,
we will often use the phrase ``equi-potential'' referring to $\Phi_{\mathrm{eff}}$.

On the other hand, the steady-state continuity equation in
the rotating frame requires,
\begin{equation}
  \nabla \cdot (\rho \vec{v}) = \rho \nabla \cdot \vec{v} + \vec{v} \cdot \nabla \rho = 0 \,.
\end{equation}
This is trivially satisfied for incompressible Riemann S-type
ellipsoids because $\nabla \cdot \vec{v}$
is zero everywhere, and the velocity vectors in the rotating frame
are always tangent to the concentric equi-potential contours
and perpendicular to the unit vector that is normal to
the ellipsoidal surface.
In our compressible models, the divergence-free velocity field
given by Eq. (\ref{vel}) will continue to be adopted,
so if the equi-potential (and iso-density) contours 
continue to be concentric ellipses \citep[as assumed by][]{LRS93}, 
the velocity vectors will always be orthogonal to the density gradient everywhere; 
this would satisfy the steady-state continuity equation exactly.  
However, as we demand that Euler's equation be satisfied 
via the solution of Eq. (\ref{bernoulli}),
the iso-density contours for our compressible models
are not guaranteed to be self-similar ellipses.
Therefore, in our current numerical method, 
the extent to which the steady-state continuity equation 
is satisfied will depend on 
how far the density shells deviate from self-similar ellipses.
For the remainder of this paper, we will only focus on
obtaining a solution of Bernoulli's function given by Eq. (\ref{bernoulli}).
We will return to issues related to the steady-state continuity equation 
in section \ref{discussion}.

\begin{figure}[ht]
\epsscale{0.8} \plotone{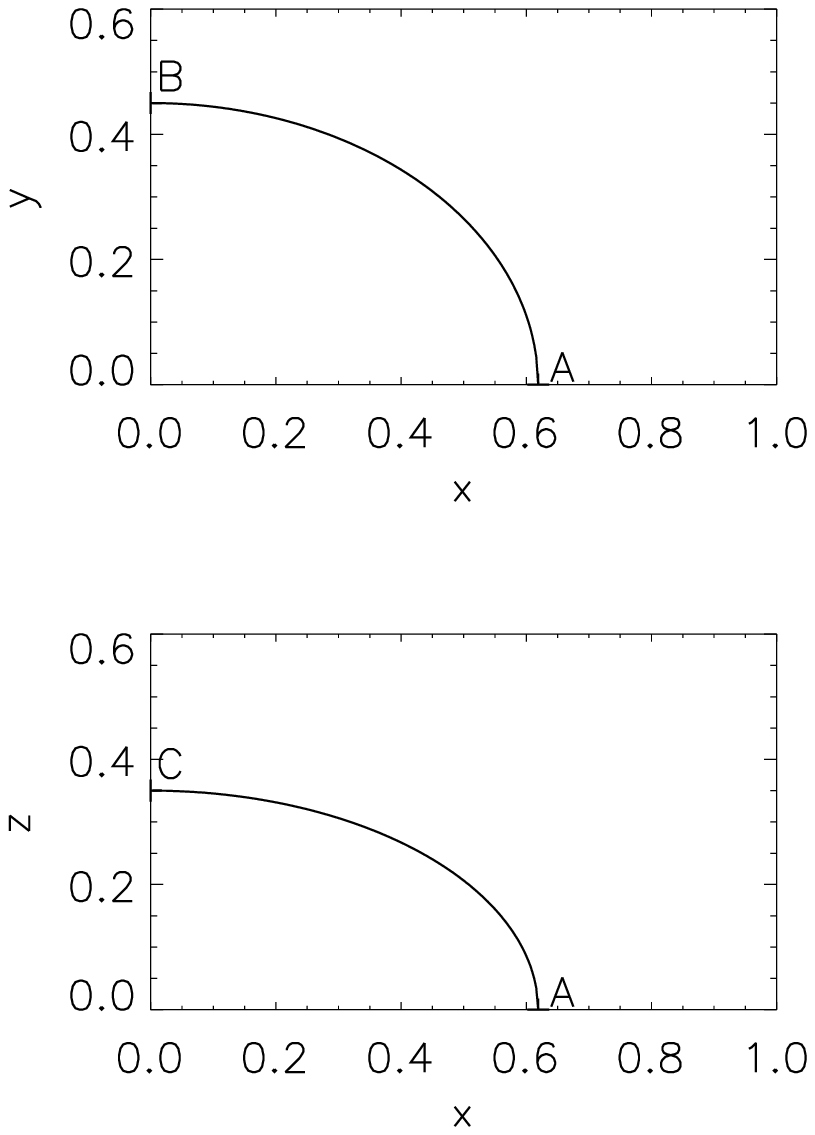}
\caption{ Three boundary conditions at the surfaces of an ellipsoid.
    Solid lines are part of the elliptical surface.   \label{boundary}}
\end{figure}

For a given choice of axes ($a$, $b$, $c$) and
a mass distribution $\rho(\vec{x})$ 
(hence $\Phi(\vec{x})$ is determined by Poisson's equation),
there are three unknown constants in Eq. (\ref{bernoulli}):
 $C_1$, $\omega$ and $\lambda$.
On the other hand, we have three boundary conditions on the three axes at the surface
of the star, i.e., points A, B and C as illustrated in Figure \ref{boundary}, 
where $H$ drops to zero.
(Note that points A, B and C are located on $X$, $Y$, and $Z$ axes 
 at $x=a$, $y=b$, and $z=c$, respectively.)
These boundary conditions give the following equations:
\begin{eqnarray}
  \Phi_A-\frac{1}{2}a^2(\omega^2+\lambda^2)+\omega \lambda ab=C_1 \label{bc1} \,,\\
  \Phi_B-\frac{1}{2}b^2(\omega^2+\lambda^2)+\omega \lambda ab=C_1 \label{bc2} \,,\\
  \Phi_C = C_1 \label{bc3} \,.
\end{eqnarray}
Subtracting equation (\ref{bc2}) from (\ref{bc1}), we obtain
\begin{equation}
  \omega^2 + \lambda^2 = \frac{2(\Phi_A-\Phi_B)}{a^2-b^2} \label{ch1} \,.
\end{equation}
Adding equation (\ref{bc2}) and (\ref{bc1}) together, we obtain
\begin{equation}
  \Phi_A + \Phi_B - 2 \Phi_C = \frac{1}{2} [a^2(\omega^2+\lambda^2)
      + b^2(\omega^2+\lambda^2) - 4 \omega \lambda a b ] \label{ch2} \,.
\end{equation}
Since the right-hand-side of this last expression is non-negative, it follows that 
$\Phi_A + \Phi_B - 2\Phi_C \geq 0$. 
It can be shown that equations (\ref{ch1}) and (\ref{ch2}) are equivalent to 
equations (29) and (30) in Chapter 7 of \cite{Ch69}.
The solutions of equations (\ref{ch1}) and (\ref{ch2}) for a direct configuration are:  
\begin{eqnarray}
  \omega^2 = (M+\sqrt{M^2-4N^2})/2 \label{omega}  \,,\\
  \lambda^2 = (M-\sqrt{M^2-4N^2})/2 \label{lambda} \,,
\end{eqnarray}
where 
\begin{eqnarray}
  M&\equiv&2(\Phi_A-\Phi_B)/(a^2-b^2) \,, \nonumber\\
  N&\equiv&\frac{(\Phi_A+\Phi_B-2\Phi_C)-\frac{a^2+b^2}{a^2-b^2}(\Phi_A-\Phi_B)}{2ab} \,. \nonumber 
\end{eqnarray}
For adjoint configurations, we only need to interchange
$\omega^2$ and $\lambda^2$.
The signs of $\omega$ and $\lambda$ are determined by the sign of $f$
(which is predetermined by the specified axis ratios), 
through Eq. (\ref{eqn:f}) and the relationship, 
\begin{equation}
   \zeta = - \lambda (b/a + a/b)  \,.
\end{equation}
With these constants and the Bernoulli's function in hand,
we can solve for $H$ throughout the interior of the configuration.

In the incompressible case ($n=0$), we know from the classical analytical 
results that the adopted velocity field generates equilibrium structures
that are uniform density ellipsoids. 
Hence, we can obtain Riemann S-type ellipsoids directly
by setting up a mass distribution that is confined by the ellipsoidal surface
defined by three axes $a$, $b$ and $c$;
then we solve Poisson's equation to get the potential everywhere;
finally, $\omega$ and $\lambda$ can be determined by equations (\ref{omega})
and (\ref{lambda}).

Compressible polytropic models ($n>0$)
having the same flow pattern as that of Riemann ellipsoids
(given by Eq. \ref{vel}), can also be 
constructed by an iterative procedure that is very similar to
 Hachisu's SCF method \citep{H86A}. 
The main steps of this method include:
(i) set up a trial ellipsoidal mass distribution defined by the choice
of $a$, $b$, and $c$, then solve Poisson's equation to obtain
the gravitational potential everywhere \citep[the details of our Poisson
solver are discussed in][]{CT99};
(ii) calculate $C_1$, $\omega$ and $\lambda$ using the three boundary
conditions as discussed above; 
(iii) calculate the enthalpy everywhere inside the configuration
 using Eq. (\ref{bernoulli});
(iv) calculate the new ``trial" density distribution according to the relationship
 between the density and enthalpy for a polytrope.
These steps are repeated until the model converges.
%%Notice that the continuity equation does not set in anywhere in our method, 
%%it is not guaranteed that our compressible models will satisfy
%%the continuity equation exactly (see later discussion).

\section{Results}
%%\subsection{Results of 3D Models}
In this section, we show our results for incompressible Riemann S-type ellipsoids
and their quasi-equilibrium, compressible counterparts. 
Since the available parameter space occupies
a large portion of Figure \ref{triaxial}, we choose to build models
only along the vertical dashed lines, which largely represent
most of the configurations.

Our models are constructed on a cylindrical grid with a resolution
of $66 \times 102 \times 128$ in the $\varpi$ (cylindrical radius), $z$
(vertical), and $\phi$ (azimuthal) directions, respectively.
Following \cite{H86A}, we adopt a set of polytropic units in which
the gravitational constant $G$, the radius of the entire grid 
$\varpi_{\mathrm{grid}}$,
and maximum density $\rho_{\mathrm{max}}$ are all set to 1.
For all of our models, we set the semiaxis $a$ to $0.619$.
We choose to present our results in such a way that readers can readily
normalize our data to any unit system they are familiar with for the 
purpose of comparison.
Table \ref{n0} shows the results for direct configurations of
incompressible ($n=0$) Riemann S-type ellipsoids.
Tables \ref{n05}, \ref{n10}, and \ref{n15} show the same results but for 
compressible counterparts of Riemann ellipsoids 
with $n=0.5$, $n=1.0$, and $n=1.5$, respectively.
Tables \ref{n0a}, \ref{n05a}, and \ref{n10a} show results for adjoint configurations 
with $n=0.0$, $n=0.5$ and $n=1.0$, respectively.

In these tables, $T$ is the rotational energy of the system, 
$M_{\mathrm{tot}}$ is the total mass, 
$J_{\mathrm{tot}}$ is the total angular momentum, 
$\rho_{\mathrm{mean}}$ is the mean density,
$W$ is the gravitational potential energy, and $S$ is the thermal energy.
The quantity $(S+T)/|W|$ measures how well our models are in virial equilibrium;
ideally, this ratio should have a value of 0.5. All of our models have values that
are very close to 0.5, especially for compressible configurations.
In Table \ref{n0},
the parameters $\omega_{\mathrm{a}}$ and $\lambda_{\mathrm{a}}$
 are analytically computed results 
for $\omega$ and $\lambda$, respectively.

Several properties of our models are worth discussing.
First, for direct configurations with a fixed value of $b/a$, 
as one proceeds from larger values of $c/a$
to smaller values, $\lambda$ starts from a positive value, then continuously
decreases; it reaches zero when the rigidly rotating Jacobi sequence is hit, 
and then becomes negative after passing the Jacobi sequence. 
The parameter $f$ also
switches sign when crossing the Jacobi/Dedekind sequences.
For adjoint configurations, a similar behavior is observed for $\omega$,
while $f$ approaches $\pm \infty$ around the Dedekind sequence
because $\omega=0$.

Second, in the parameter space above the Jacobi/Dedekind sequence 
where $f<0$, $\omega$ is positive for direct configurations, and negative
for adjoint configurations, 
which reflects the fact that the Jacobi mode in 
Maclaurin spheroids is the ``forward moving'' wave but the Dedekind mode
is the ``backward moving'' wave.
This also implies that in the adjoint configurations with $f<0$,
the ellipsoidal pattern is moving in a direction that is
opposite to the overall rotation!
Furthermore, when the critical Jacobi/Dedekind sequence is passed, $f$ switches sign; 
so does $\lambda$ in direct configurations and $\omega$ in adjoint configurations.
In Maclaurin spheroids, this is where the frequency of the backward moving
wave becomes positive in the inertial frame; 
hence, it is where the secular bar-mode
instability sets in through gravitational radiation
(Chandrasekhar-Friedman-Schutz instability) or viscosity \citep{Ch69, FS78}.

Finally, for softer equations of state, the SCF technique failed to 
converge to models with relatively small values of $c/a$
because, in these cases, the centrifugal force at the boundary point A
(see Figure \ref{boundary}) would exceed the gravitational force.
%%resulting mass to shrink from equator along the longest axis. 
%%We attribute this to the mass shedding limit at the equator.
In fact, for $n=1$ and $n=1.5$, even the Jacobi sequence is not reached.
This is consistent with previous studies that have indicated that
the Jacobi sequence cannot exist for $n>0.808$ \citep{J64, HE82, LRS93}. 
Furthermore, as was pointed out by \cite{HE82} in their analysis of 
uniformly rotating polytropes, 
the mass-shedding limit never extends much beyond the bifurcation point
(where the Jacobi/Dedekind sequence branches off the Maclaurin spheriod sequence)
 even for very small $\mathrm{n}$:
for $0.1 \leq \mathrm{n} \leq 0.5$, this mass-shedding limit corresponds to 
$0.14 \leq T/|W| \leq 0.16$. Our $\mathrm{n}=0.5$ models are
consistent with their results.
Therefore, the available parameter space for compressible counterparts
of Riemann ellipsoids becomes smaller as the equation of state
becomes softer. 

\begin{figure}[ht]
\epsscale{1.0} \plotone{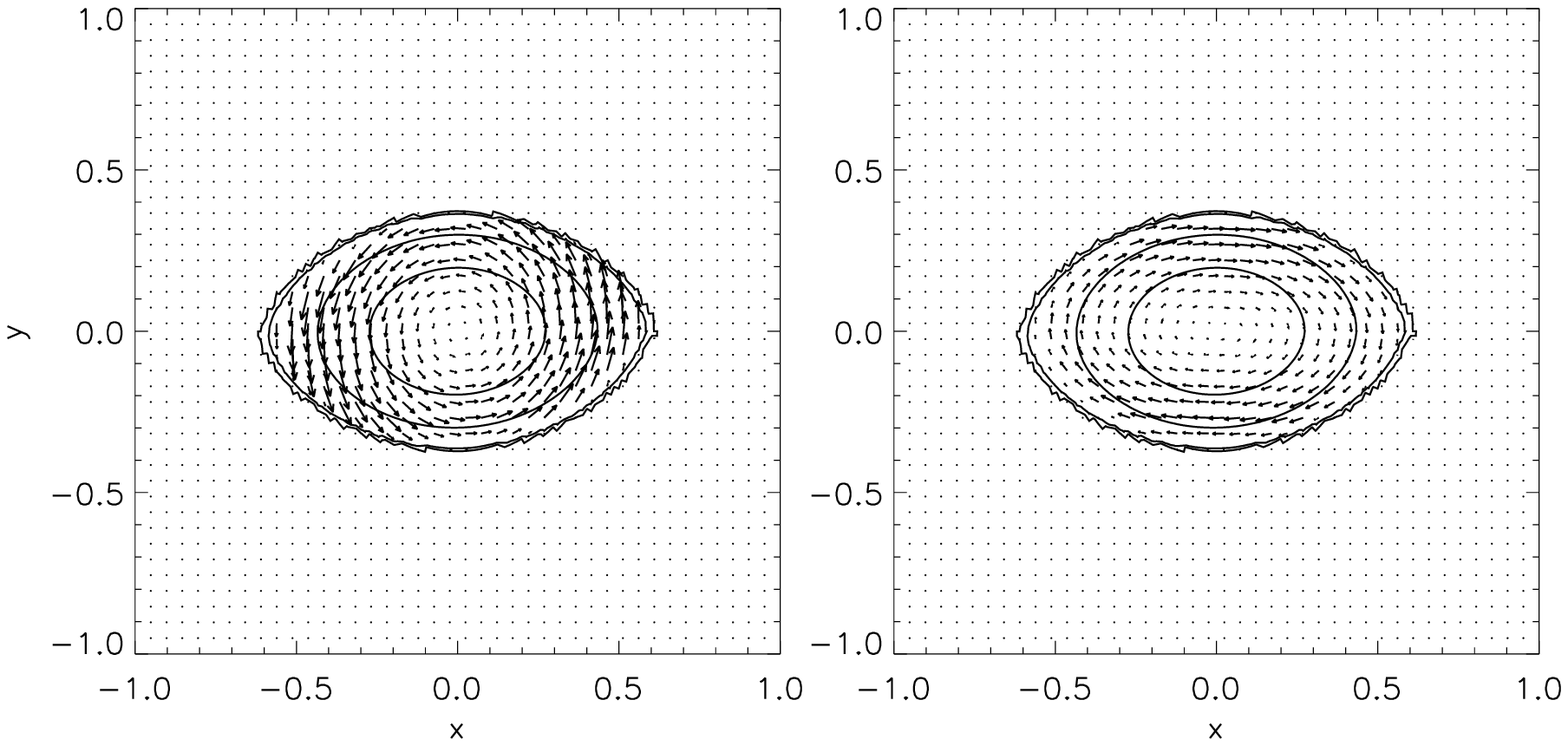}
\caption{ Equatorial iso-density contours and velocity fields 
    in the inertial frame (left) and rotating frame (right) 
    for an $n=0.5$, direct configuration with $b/a=0.59$ and $c/a=0.462$. 
    Density contours correspond to
    $\rho/\rho_{max} = 0.01, 0.1, 0.5, 0.8$ from the outermost shell
    to innermost shell. 
    This model has $\omega=0.905$ and $\lambda=0.0754$,
    which corresponds to a vorticity $\zeta=-0.17$ in the rotating frame.
    So the ellipsoidal configuration is spinning prograde (in the same
    direction as the overall fluid rotation, i.e., counterclockwise)
    rapidly in the inertial frame, but the fluid is moving more slowly
    and retrograde as viewed from the rotating frame.
    \label{direct05}}
\end{figure}

\begin{figure}[t]
\epsscale{1.0} \plotone{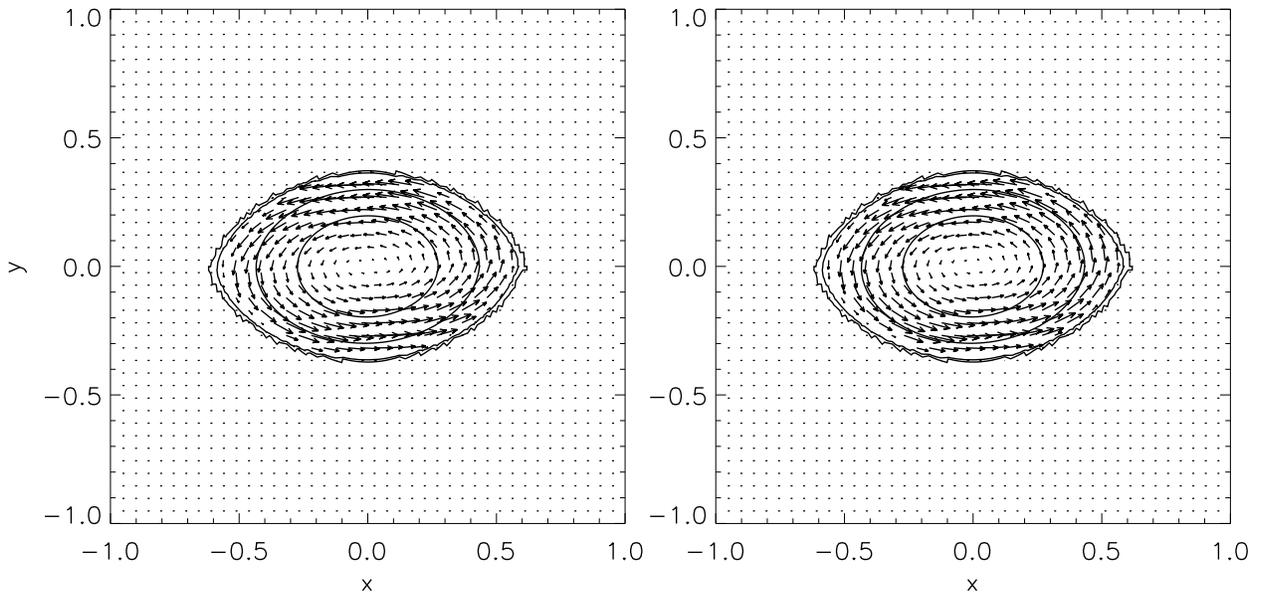}
\caption{ The same as Fig. \ref{direct05} but for the adjoint configuration.
    This model has $\omega=-0.075$ and $\lambda=-0.905$,
    which corresponds to $\zeta=2.07$,
    so the ellipsoidal configuration has a slow retrograde (clockwise) spin
    in the inertial frame (left), but the fluid is moving rapidly prograde
    as viewed from the rotating frame (right).
         \label{adjoint05}}
\end{figure}

\begin{figure}[ht]
\epsscale{1.0} \plotone{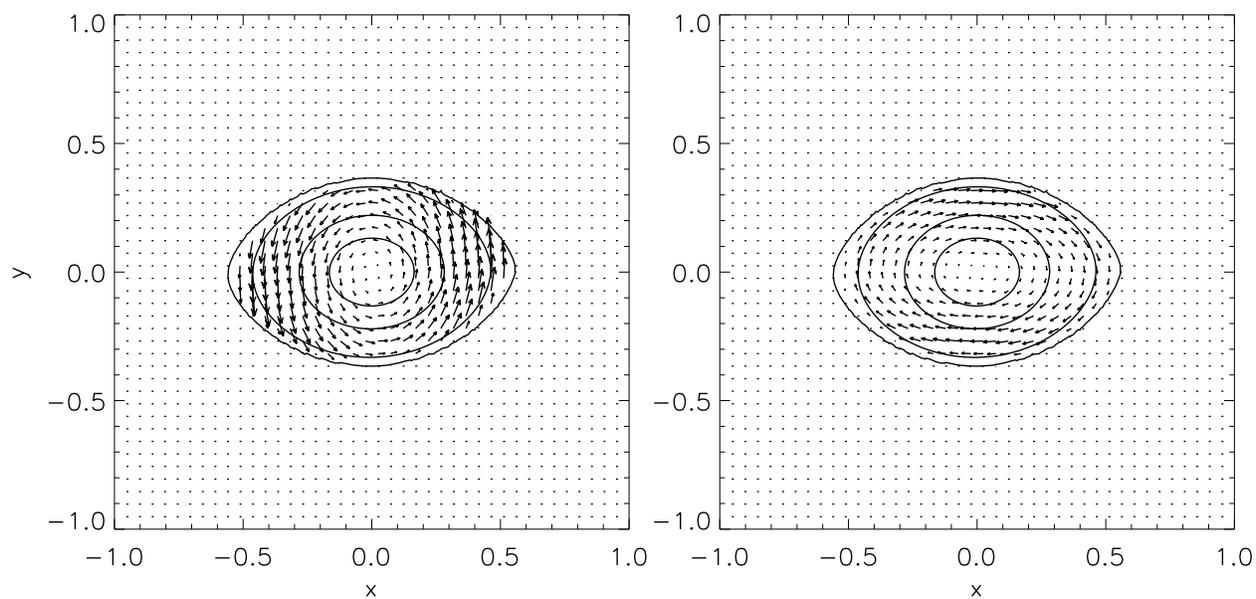}
\caption{  The same as Fig. \ref{direct05} but
    for an $n=1$, direct configuration with $b/a=0.59$ and $c/a=0.564$. 
    Density contours correspond to 
    $\rho/\rho_{max} = 0.01, 0.1, 0.5, 0.8$ from the outermost shell
    to innermost shell. 
    This model has $\omega=0.743$ and $\lambda=0.2011$,
    which corresponds to $\zeta=-0.46$.
    So the ellipsoidal configuration is spinning prograde (counterclockwise)
    rapidly in the inertial frame (left),
    but the fluid is moving retrograde (clockwise) in the rotating frame (right).
    \label{direct}}
\end{figure}

\begin{figure}[t]
\epsscale{1.0} \plotone{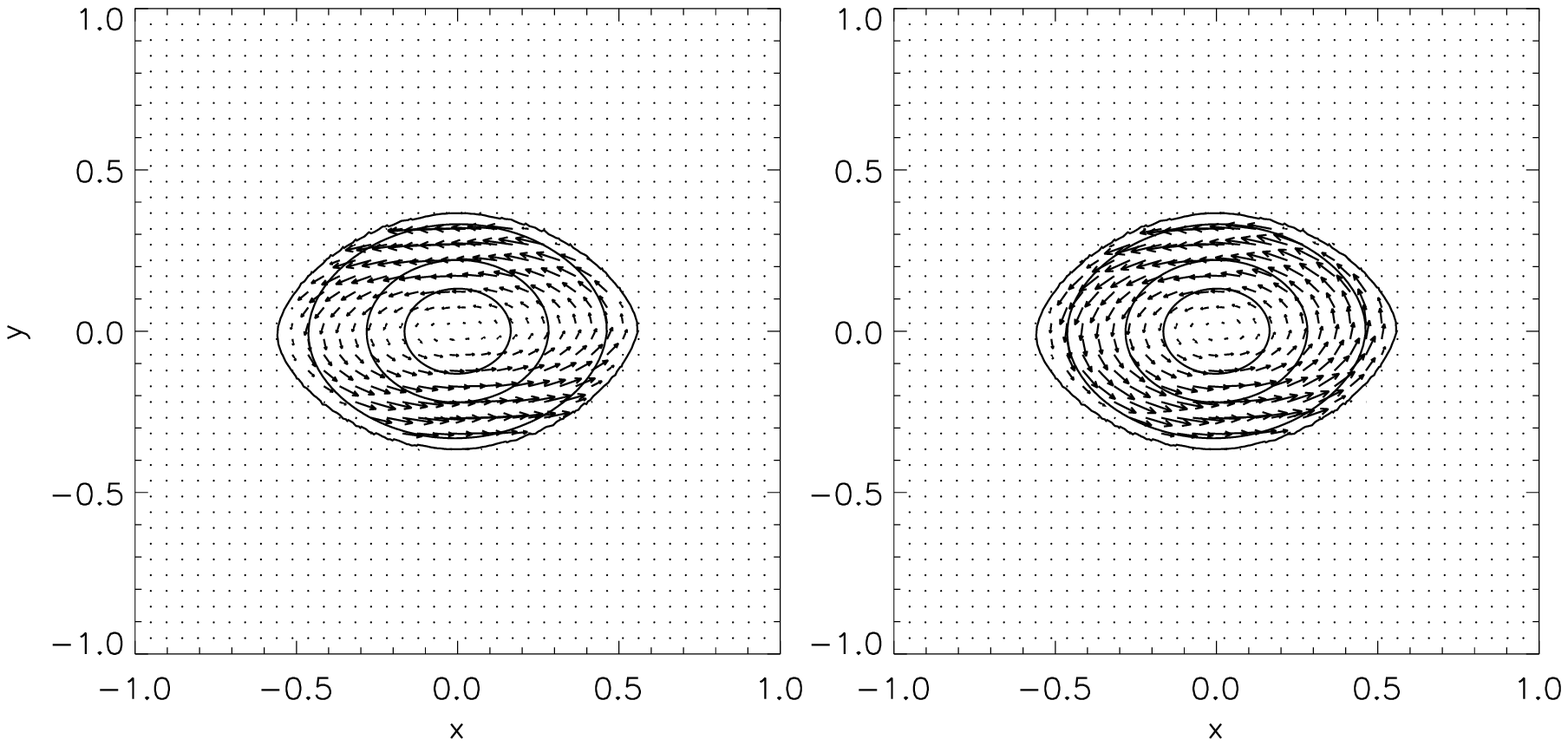}
\caption{ The same as Fig. \ref{direct} but for the adjoint configuration. 
    This $n=1$ model has $\omega=-0.201$ and $\lambda=-0.7433$,
    which corresponds to $\zeta=1.70$.
    So the ellipsoidal configuration is spinning retrograde (clockwise)
    in the inertial frame,
    but the fluid is moving prograde (counterclockwise) rapidly in the rotating frame.
         \label{adjoint}}
\end{figure}

Figures \ref{direct05} and \ref{adjoint05}
show equatorial-plane iso-density contours and velocity fields
 for direct and adjoint configurations
of our $n=0.5$ model with $b/a=0.59$ and $c/a=0.462$. 
Both the rotating frame velocity field $\vec{v}$ and inertial frame velocity field
 $\vec{u} = \vec{v}+\vec{\omega} \times \vec{r}$ are plotted for comparison.
The flow patterns in
these two models are different from each other although they have identical geometrical shapes.
It is also clear that the fluid motion in the adjoint configuration is dominated by 
internal motion rather than rotation. 
As another illustration, Figures \ref{direct} and \ref{adjoint} illustrate
the structure of our $n=1$ model with $b/a=0.59$ and $c/a=0.564$.
Notice that $\omega$, which is also the pattern speed of the ellipsoid,
 is negative in both Fig. \ref{adjoint05} and Fig. \ref{adjoint}, 
so the ellipsoidal figures in both plots
are spinning retrograde (clockwise) relative to the overall fluid rotation.

Comparisons between previous published results and our models 
yield very good agreement.
For example, from Table IV in Chapter 6 of \cite{Ch69}, 
a Jacobi ellipsoid with $\lambda=0$ and $b/a$=0.60, $c/a=0.434$,
has $\omega^2/(\pi G \rho)=0.3373$ 
and $J_{\mathrm{tot}}/\sqrt{G M_{\mathrm{tot}}^3 \bar{a} }= 0.3356$,
where $\bar{a}=(abc)^{\frac{1}{3}}$.
The closest match among our models shown in Table \ref{n0}, with $b/a=0.59$, $c/a=0.436$,
has $\omega^2/(\pi G \rho)=0.3370$, 
$J_{\mathrm{tot}}/\sqrt{(G M_{\mathrm{tot}}^3 \bar{a} }= 0.3321$,
and an almost vanishing $\lambda=0.0108$, which shows that this model 
from our Table is very close to the Jacobi sequence. 
It should be noticed that, due to the discrete nature of our grid,
we cannot construct models with axis ratios that will place the models precisely
on the Jacobi/Dedekind sequence,
but we can construct models that are very close to this sequence
(with very small values of $\lambda$ or $\omega$).
As another comparison, the last four columns in Table \ref{n0} 
give analytically computed results of $\omega_{a}$ and $\lambda_{a}$ for 
Riemann S-type ellipsoids having the specified axes ratios $b/a$ and $c/a$,
as well as the ratio between numerical and analytical results
(only models with $a < b < c$ have been computed).
To obtain these analytical results,
we used standard incomplete elliptic integrals to evaluate the 
potential field of a homogeneous ellipsoid 
\cite[see their Table 2-2]{BT87}.
In virtually all cases, the error in $\omega$ is $\lesssim$ a few percent.
On the other hand, for the direct configurations
the fractional error in $\lambda$ can be as large as 50\%;
the error is largest for models whose $\lambda$ almost vanishes.
This is understandable because a small variation in the magnitude of $\lambda$
will result in a large percentage difference when $\lambda$ is approaching zero.
Also some mismatch between numerical and analytical results
may arise because the discrete nature of our grids makes it difficult
for us to define an exact axis ratio for our models.

For compressible equations of state, we can compare our results with those
presented by \cite{LRS93}, who used an energy variational method 
to determine approximate equilibrium configurations 
for ellipsoidal self-gravitating systems. 
\cite{LRS93} assumed that the iso-density contours of
each configuration are self-similar concentric ellipses, 
which as we have already emphasized is not the case.
In their Table 4, for compressible analogues of Jacobi ellipsoids, 
a configuration with $b/a=0.75$ and $c/a=0.4983$ has $T/|W|=0.1407$;
the closest match among our $n=0.5$ models (see Table \ref{n05}) 
with $b/a=0.74$ and $c/a=0.487$ has $T/|W|=0.135$. 
Other quantities also match well, after a proper normalization 
is applied as prescribed by their equation (3.27).
We omit more comparisons here and leave the rest to interested readers. 

Other models that are of interest are the $f=-2$ irrotational sequence
\citep{UE98}.
It turns out that the $f$ parameter is very sensitive to the axis ratios, 
so we could not build models that are very close to the irrotational sequence
due to the discrete nature of our computational grids.
Our model that comes closest has $f=-1.9$;
it is a direct configuration with $n=0.5$ and its other parameters
are listed in Table \ref{n05}.

%%\subsection{Results from 3D Hydrodynamical Evolutions}

%%In order to see if these models are indeed stable equilibrium configurations,
%%we need to use 3D hydrodynamics technique to evolve them. But for practical 
%%reason, we can't require evolutions for all the models, which far more beyond
%%our computational resources. On the other hand, it is also impossible to evolve
%%an $n=0$ incompressible model, because the sound speed goes to infinity in this case. 
%%We decided to evolve two models, one of which is very close to the Jacobi sequence,
%%and the other one is very close to the Dedekind sequence.

\section{Discussion and Future Work}
\label{discussion}
We have presented a new method to construct 3D models for Riemann S-type ellipsoids ($n=0$),
and the method has been extended to construct compressible counterparts
that share the same velocity field as that of Riemann S-type ellipsoids.
With this method, we have been able to build 3D models that cover
almost the entire parameter space of Riemann S-type ellipsoids.
Our results are in good agreement with previous studies.
We expect that this method can also be straightforwardly extended to 
the construction of Roche-Riemann ellipsoids.

\begin{figure}[t]
\epsscale{1.0} 
\plotone{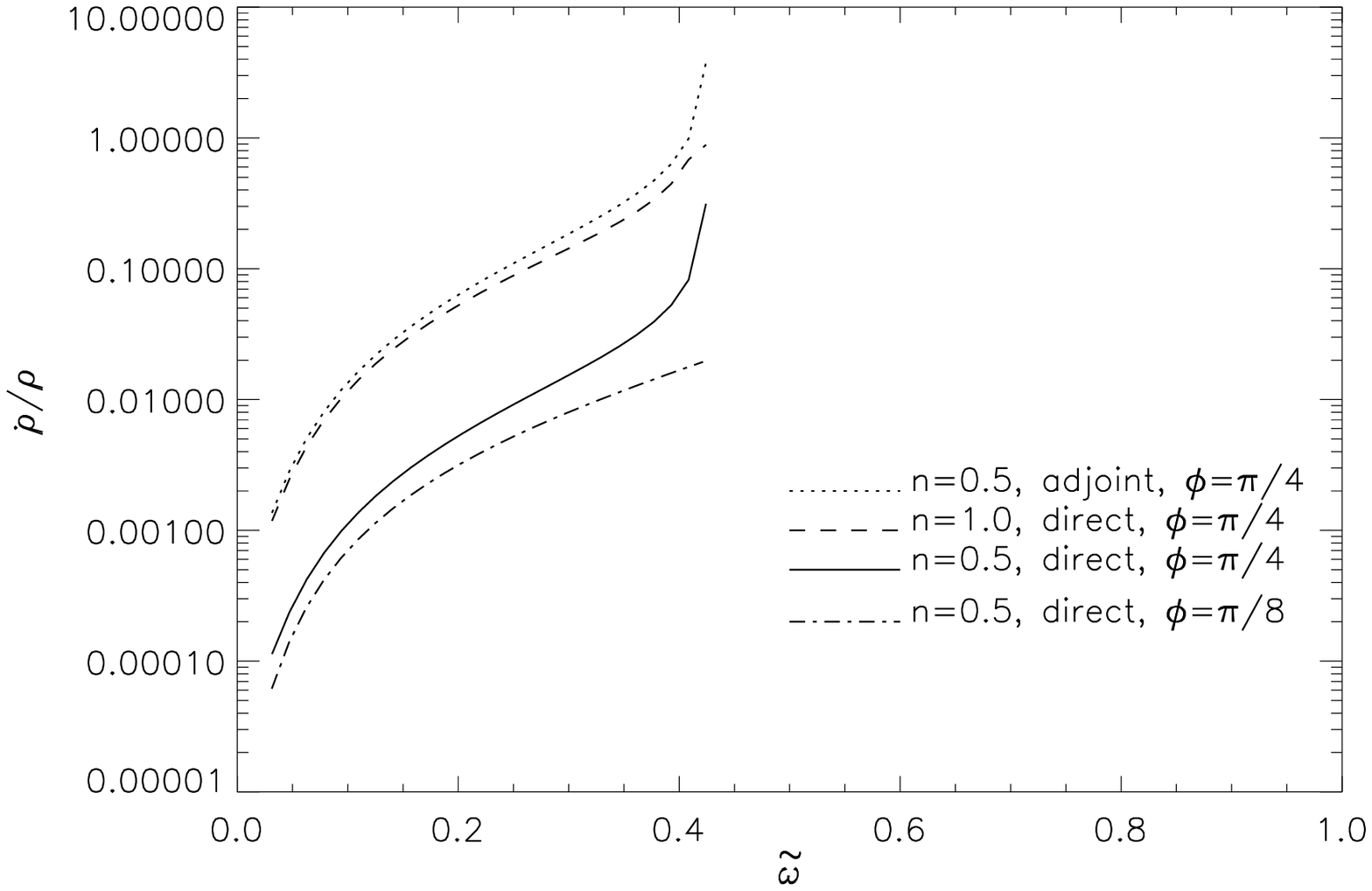}
\caption{$|\dot{\rho}|/\rho$ is plotted in the equatorial plane 
    as a function of cylindrical radius $\varpi$
    at angles $\phi=\pi/4$  and $\phi=\pi/8$ 
    for the models shown in Figs. \ref{direct05}, \ref{adjoint05}, and \ref{direct}.
    $|\dot{\rho}|/\rho$ is small in the interior of each model,
    but can become quite large near the outer edge where $\rho$ approaches zero.
    Also, $|\dot{\rho}|/\rho$ is smaller for the models with the stiffer equation of state
    and for direct configurations.
         \label{divrhov}}
\end{figure}

By design, our compressible models satisfy the steady-state Euler's equation.
However, it is not guaranteed that they simultaneously satisfy 
the steady-state continuity equation exactly,
because this additional constraint was not built into our method. 
%%The steady-state continuity equation requires, 
%%\begin{equation}
%%  \nabla \cdot (\rho \vec{v}) = \rho \nabla \cdot \vec{v} + \vec{v} \cdot \nabla \rho = 0 \,,
%%\end{equation}
%%This is trivially satisfied in the incompressible cases because $\nabla \cdot \vec{v}$
%%is zero everywhere, and the velocity vectors in the rotating frame
%%are always tangent to the self-similar equi-potential contours.
%%In \cite{LRS93},
%%the iso-density contours (equi-potential contours) are assumed 
%%to be self-similar concentric ellipses,
%%which is precisely true for incompressible Riemann S-type ellipsoids
%%and helps to satisfy the equation of continuity. 
%%Here we will relax this condition.
%%In compressible cases, if the equi-potential (iso-density) contours remain to be self-similar
%%concentric ellipses, these two conditions still keep the continuity equation
%%to be exact.  
%%However, the latter condition is no longer true in compressible cases. 
Because the adopted velocity flow-field $\vec{v}$ in our models is divergence-free,
the violations of the steady-state continuity equation will only arise 
to the extent that the iso-density contours deviate from concentric ellipses.
As we can see from Figures \ref{direct05} - \ref{adjoint},
the outer low-density shell in compressible models 
becomes somewhat pointed along the major axis.
However, the deviation from concentric ellipses is small and almost negligible
in the inner, high-density regions. 
Furthermore,  a comparison between Fig. \ref{direct05} and Fig. \ref{direct}
shows that this deviation becomes smaller in less compressible models.
In order to quantitatively measure the violation of the steady-state 
continuity equation, we calculated 
$|\dot{\rho}|/\rho \equiv |\nabla\cdot(\rho\vec{v})|/\rho$ 
for the models shown in Figs \ref{direct05}, \ref{adjoint05},
and \ref{direct}. 
As long as $|\dot{\rho}|/\rho << \sqrt{\pi G \rho_{max}} = \sqrt{\pi}$ for a given configuration,
we can consider that the model has a reasonablly good quasi-equilibrium structure.
In Figure \ref{divrhov}, we have plotted $|\dot{\rho}|/\rho$
in the equatorial plane as a function of cylindrical radius $\varpi$
along different azimuthal angles, $\phi=\pi/8$ and $\phi=\pi/4$.
In fact, this quantity remains small ($\lesssim 0.01$ for $n=0.5$ 
and $\lesssim 0.1$ for $n=1.0$) throughout the majority of the configuration,
but not surprisingly near the surface it asymptotically approaches infinity
because $\rho$ goes to zero.
As the figure illustrates, $|\dot{\rho}|/\rho$ is smaller for the stiffer equation of state.
We also note that $|\dot{\rho}|/\rho$ is nearly zero along the X and Y axes
because the velocity vectors in the rotating frame
 along the X (Y) axis only have a nonzero Y (X) component
and are therefore perpendicular to the density gradient there.
Thus, we expect our compressible models to represent fairly 
good quasi-equilibrium configurations,
especially in nearly incompressible cases.

\begin{figure}[t]
\epsscale{1.0}
%%\plotone{div.direct.ps}
\plotone{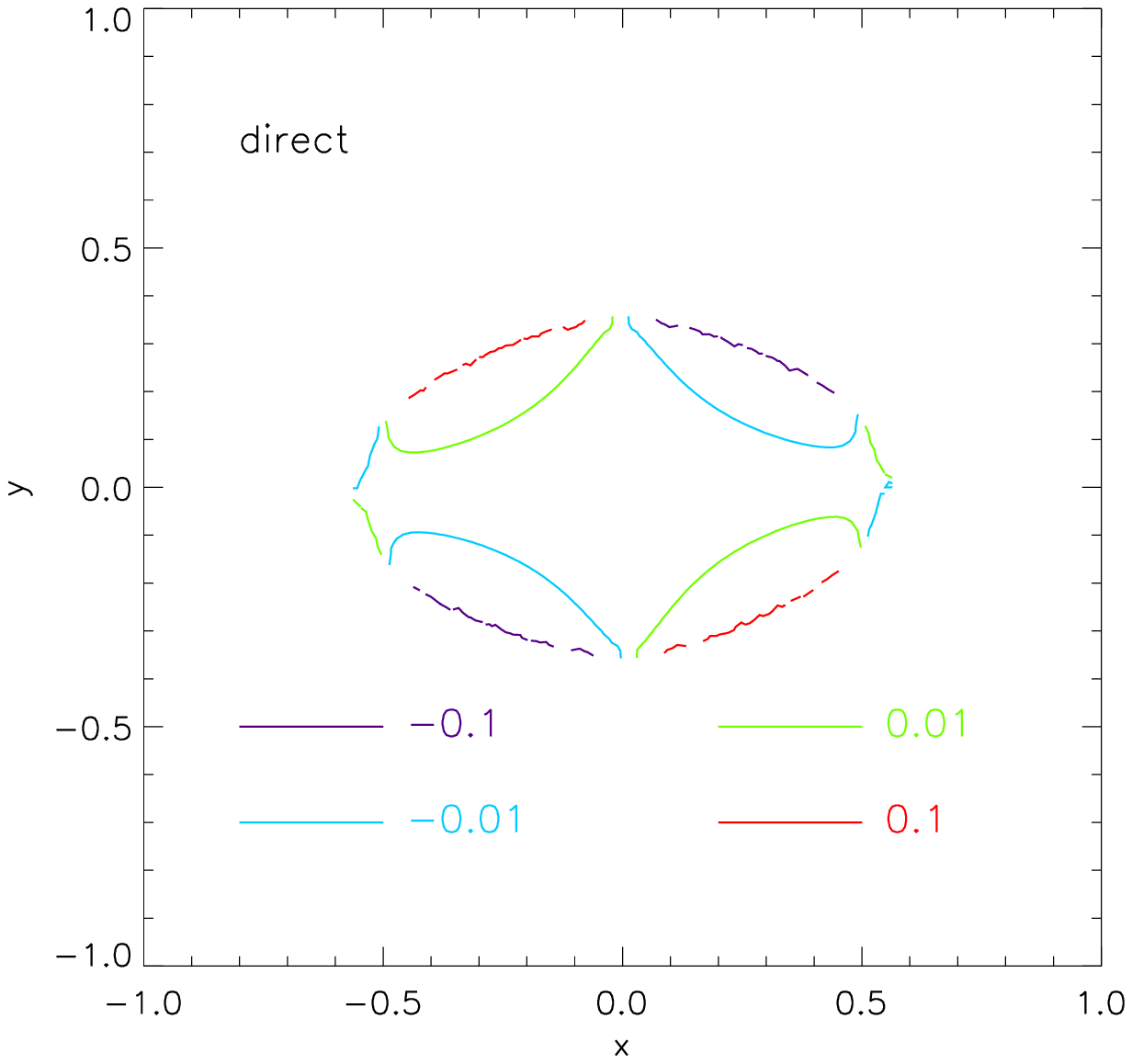}
\caption{Contour map of $\nabla\cdot(\rho\vec{v})/\rho$ is plotted in the equatorial plane
    for the direct model shown in Fig. \ref{direct05}.
         \label{divrhov2}}
\end{figure}

\begin{figure}[t]
\epsscale{1.0}
\plotone{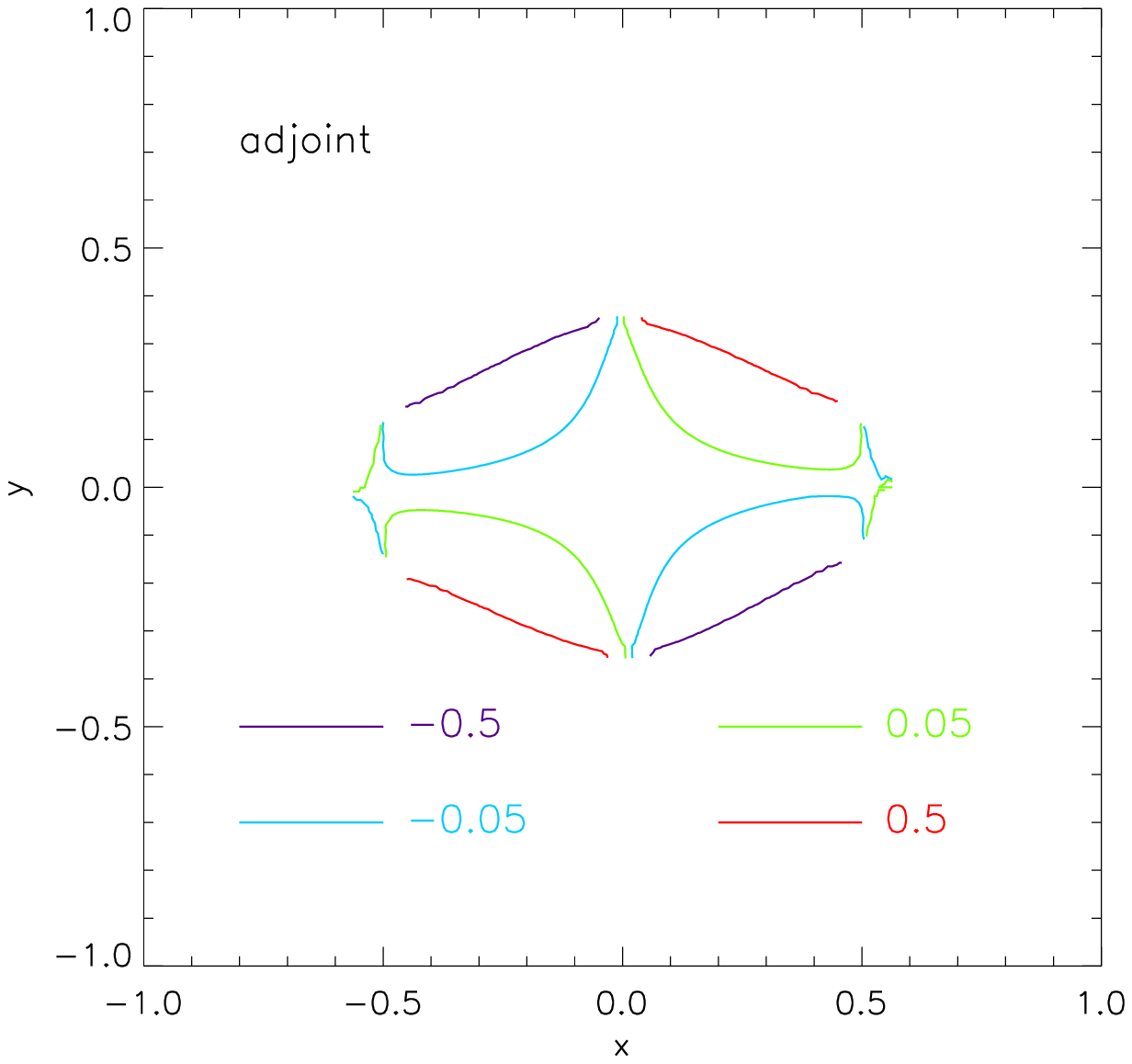}
\caption{Contour map of $\nabla\cdot(\rho\vec{v})/\rho$ is plotted in the equatorial plane
    for the adjoint model shown in Fig. \ref{adjoint05}.
         \label{divrhov3}}
\end{figure}

\begin{figure}[t]
\epsscale{1.0}
\plotone{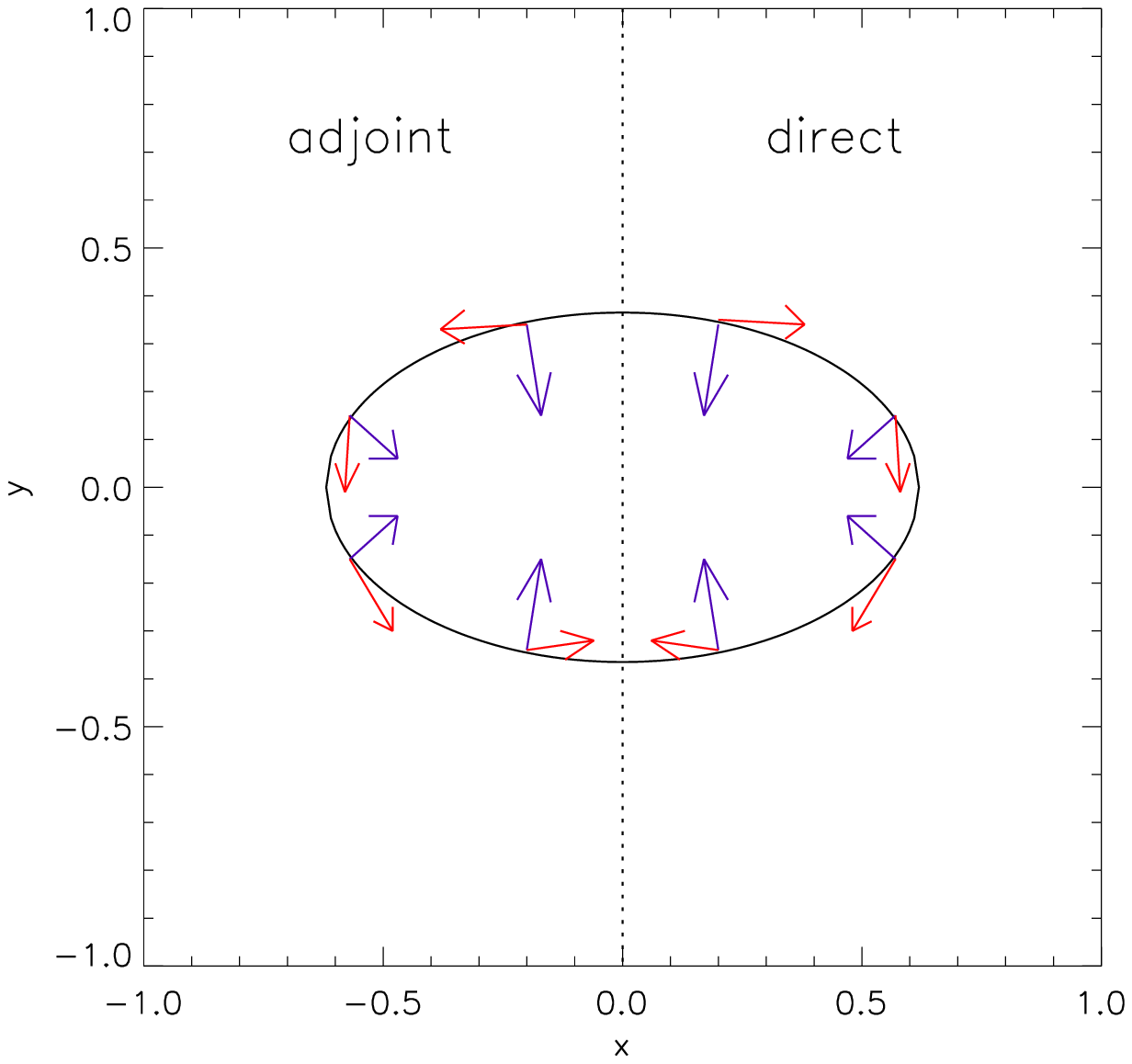}
\caption{A combined schematic drawing of the density gradient (purple vectors) and
    velocity vectors (red) in the rotating frame for the direct (right panel)
    and adjoint (left panel)
    configurations shown in Fig. \ref{direct05} and \ref{adjoint05}. 
    The black ellipse denotes the surface of the ellipsoidal model.
    Notice that the fluid is moving prograde (retrograde) 
    in the rotating frame for the chosen direct (adjoint) configuration.
         \label{shape}}
\end{figure}

On the other hand, this point was brought to our attention by an anonymous referee:
the relatively large violation of the steady-state continuity equation
on the very surface layer can potentially change the shape of
the surface in hydrodynamical evolutions.
To understand how the shape of the surface might be changed by the flow field,
it is important to know the sign of $\nabla\cdot(\rho\vec{v})/\rho$ 
along the surface.
Since we are adopting a divergence free velocity field,
the sign of $\nabla\cdot(\rho\vec{v})/\rho$ is determined by
the sign of $\vec{v}\cdot\nabla\rho$,
which also measures to what degree the velocity vectors are orthogonal to
density gradient.
If the velocity vector points slightly outward (inward) from the surface,
$\vec{v}\cdot\nabla\rho$ will be negative (positive),
because the density gradient is negative outward.
Figs. \ref{divrhov2} and \ref{divrhov3} show contour maps of 
$\nabla\cdot(\rho\vec{v})/\rho$ in the equatorial planes of 
the direct and adjoint models
shown in Figs. \ref{direct05} and \ref{adjoint05}, respectively.
(We note that the contour pattern for direct configurations will switch
from Fig. \ref{divrhov2} to Fig. \ref{divrhov3} if $\lambda$ flips sign.)
A $\pi$ symmetry is observed for both patterns.
We also note that, in each quadrant, $\nabla\cdot(\rho\vec{v})/\rho$ flips sign,
which implies that the direction of velocity vectors will change 
from inward (outward) to outward (inward) from the surface within one single quadrant.
Accordingly, Fig. \ref{shape} shows a combined schematic drawing of 
how the velocity vectors in the rotating frames are aligned with the surface 
for these two patterns
(the panel on the left is for the adjoint figure shown in Fig. \ref{adjoint05},
the panel on the right is for the direct figure shown in Fig. \ref{direct05}).
In this figure, the black ellipse denotes the actual surface
(which should not be a perfect ellipse),
red arrows denote velocity vectors on the surface,
purple arrows denote density gradient vectors,
which are drawn pointing inward because of their negative sign.
For direct configurations, in which the figure rotation is larger than the internal motion,
these mismatches (nonvanishing $\vec{v}\cdot\nabla\rho$) on top of a large rotation 
are likely to generate spiral arms;
whereas, for adjoint configurations, the internal motion is larger than figure rotation,
so these mismatches tend to generate waves along the surface.

Using the 3D hydrodynamics code developed by the LSU group \citep{MTF02},
we have carried out nonlinear 3D hydrodynamical evolutions for 
a subset of the nearly incompressible models discussed here in Table \ref{n05}.
Despite the fact that the models do not fully satisfy the steady-state 
continuity equation, they are in very good force-balance initially
and, as predicted by the above discussion, the models are fairly long-lived, 
holding their initial ellipsoidal shapes over the time that is followed 
by our simulations ($\sim$ 15 - 20 dynamical times).
We also observed the formation of narrow spiral arms (low amplitude surface waves) 
containing low density materials in direct (adjoint) configurations.
The results of these hydrodynamical evolutions will be presented in a separate paper
in the context of our ongoing study of the stability of
Riemann S-type ellipsoids.
Since our ellipsoidal models are fairly long-lived and 
have exactly the same elliptical flow field as Riemann S-type ellipsoids,
they should provide good initial states for a nonlinear, hydrodynamical 
study of dynamical instabilities in
Riemann S-type ellipsoids.
We especially have in mind the elliptical strain instability
that is associated with elliptical stream lines,
which should set in on a dynamical time scale.
Through such a study,
we hope to provide insight into the stability of ellipsoidal
configurations that will complement the work of \cite{LL96}.
This should also help answer whether the evolutionary path of a GRR-driven
star toward the Dedekind sequence is viable.

%% use either \url or \anchor, with the HREF as the first argument and the
%% text to be printed in the second.

\acknowledgments

I would like to thank Joel Tohline for valuable suggestions, discussions
and carefully reading through this paper. 
I also thank Juhan Frank and Patrick Motl for helpful comments.
It is a pleasure to thank Norman Lebovitz for drawing our attention
to the particular analytical expression of the velocity field of 
Riemann S-type ellipsoids used in this study.
I would also thank Howard Cohl for developing the Poisson solver
that was an important ingredient in this work.
I thank the anonymous referee for insightful suggestions
to analysis our models.
This work was partially supported by NSF grants AST-0407070 and PHY-0326311,
and by the Center for Computation \& Technology at LSU.
\begin{rotate}
%\begin{landscape}
%\textheight 11in
%\topmargin -0.1in
%\headheight -0.5in
%%\footheight 0.5in
%\oddsidemargin -0.5in

\begin{deluxetable}{cccccccccccccc}
\tabletypesize{\scriptsize}
\tablecolumns{14}
\tablewidth{0pt} 
\tablecaption{Direct Configurations of Riemann S-type Ellipsoids ($n=0$)\label{n0}}
\tablehead{ 
\colhead{$b/a$} & \colhead{$c/a$} & \colhead{$\omega$} &
\colhead{$\lambda$} & \colhead{$T/|W|$} & \colhead{$J_{\mathrm{tot}}$} &
\colhead{$M_{\mathrm{tot}}$} & 
\colhead{$W$} & \colhead{$(S+T)/|W|$} & \colhead{$f$} &
\colhead{$\omega_{a}$} & \colhead{$\omega / \omega_{a}$} &
\colhead{$\lambda_{a}$} &\colhead{$ \lambda / \lambda_{a}$}
} \startdata

%%\begin{table}[]
%%\begin{center}
%%\caption{Direct Configurations of Riemann S-type Ellipsoids ($n=0$)\label{n0}}
%%\begin{tabular}{|c|c|c|c|c|c|c|c|c|c|c|}
%%  \tableline
%%   $b/a$ & $c/a$ & $\omega$ & $\lambda$ & $T/|W|$ & $J_{\mathrm{tot}}$ &
%%   $M_{\mathrm{tot}}$ & $\rho_{\mathrm{mean}}$ & $W$ & $(S+T)/|W|$ & $f$ \\ 
%%   \tableline

   0.90  &  0.923 &  1.030 &  0.7550 &  6.6e-3&  0.03183 & 0.8232& -0.6989 & 0.4989 & -1.5 & &\\
   0.90   & 0.795 & 1.144 & 0.4310 & 0.046 & 0.07018 & 0.7086 &  -0.5429 & 0.4992 & -0.76 & 1.14704 &0.997 & 0.43181 &0.998\\
   %%0.90  &  0.744  & 1.146 &  0.3279 &  0.064 & 0.07533 & 0.6638 &  -0.4857 &  0.4987& -0.58 & &\\
   %%0.90  &  0.692  & 1.150 &  0.2440 &  0.082 & 0.07751 & 0.6174 &    -0.4290 &  0.4991& -0.43 & &\\
   0.90  &  0.641  & 1.119 &  0.1394 &  0.101 & 0.07764 & 0.5720 &    -0.3759 &  0.4988& -0.25 & 1.13137& 0.989&0.15077 & 0.925\\
   0.90  &  0.590  & 1.099 &  0.0569 &  0.122 & 0.07589 & 0.5261 &    -0.3249 &  0.4987& -0.10  &1.10661 &0.993 &0.06406 & 0.888\\
   0.90  &  0.564  & 1.090 &  0.0210 &  0.132 & 0.07452 & 0.5033 &    -0.3007 &  0.4989& -0.04 & 1.09034& 0.999&0.02033&1.033\\
   0.90  &  0.538  & 1.069 & -0.0250 &  0.144 & 0.07268 & 0.4801 &    -0.2768 &  0.4991& 0.05 & 1.07148& 0.998&-0.02324 &1.075\\
   0.90  &  0.487  & 1.015 & -0.1205 &  0.167 & 0.06814 & 0.4341 &    -0.2316 &  0.4992& 0.24 & 1.02639&0.989& -0.10880 &1.108\\
   %%0.90  &  0.436  & 0.945 & -0.2188 &  0.191 & 0.06254 & 0.3886 &    -0.1900 &  0.4986& 0.47 & 0.98665&-0.19571\\
   %%0.90  &  0.385  & 0.909 & -0.2736 &  0.218 & 0.05607 & 0.3431  &   -0.1518 &  0.4987& 0.61 & &\\
   0.90  &  0.333  & 0.797 & -0.3885 &  0.246 & 0.04860 & 0.2970 &    -0.1167 &  0.4983& 0.98 & 0.79257& 1.006&-0.39224&0.990\\
   \tableline
   0.74  &  0.769  & 1.099 &  0.5604  & 0.030 & 0.03807 & 0.5682 &  -0.3744  & 0.4982 & -1.1 & &\\
   0.74  &  0.692  & 1.133 &  0.3840  & 0.058 & 0.04678 & 0.5116 &  -0.3134  & 0.4986& -0.71 &1.13215 & 1.001&0.38599&0.995\\
   %%0.74  &  0.641  & 1.132 &  0.2748  & 0.078 & 0.04908 & 0.4741  &  -0.2751  & 0.4982& -0.51 & &\\
   %%0.74  &  0.590  & 1.116 &  0.1715  & 0.099 & 0.04937 & 0.4357  &  -0.2378  & 0.4990& -0.32 &1.11990 &0.18020\\
   0.74  &  0.538  & 1.094 &  0.0774  & 0.121 & 0.04836 & 0.3982  &  -0.2034  & 0.4990& -0.15 &1.09508 & 0.999&0.08149 &0.950\\
   0.74  &  0.513  & 1.078 &  0.0303  & 0.133 & 0.04734 & 0.3791   &  -0.1857  & 0.4991& -0.06 &1.07883 & 0.999&0.03474&0.872\\
   0.74  &  0.487  & 1.054 & -0.0208  & 0.145 & 0.04594 & 0.3560   &  -0.1700  & 0.4988&  0.04 &1.05883 & 0.995 &-0.01363& 1.526\\
   0.74  &  0.436  & 1.003 & -0.1184  & 0.171 & 0.04284 & 0.3223   &  -0.1400  & 0.4987&  0.25 &1.00963  &0.993&-0.10852 &1.091\\
   %%0.74  &  0.385  & 0.944 & -0.2095  & 0.198 & 0.03864 & 0.2841   &  -0.1117  & 0.4988&  0.46 & &\\
   0.74  &  0.333  & 0.847 & -0.3230  & 0.228 & 0.03389 & 0.2463   &  -0.0863  & 0.4983&  0.80 & 0.85686& 0.988&-0.31006&1.042\\
   0.74  &  0.282  & 0.715 & -0.4523  & 0.259 & 0.02842 & 0.2081   &  -0.0634  & 0.4968&  1.3 &0.72756 & 0.983&-0.43781 &1.033\\

   \tableline

   0.59  &  0.718  & 0.948   &0.7677  & 0.014 & 0.01197 & 0.4208  &  -0.2243 & 0.4924 & -1.9 & &\\
   0.59  &   0.538 &   1.087 &   0.2338 &   0.093 &  0.02876 &  0.3157  &    -0.1374 &  0.4982& -0.49 &1.08732 & 0.999 &0.23529& 0.994\\
   %%0.59  &   0.487 &   1.063 &   0.1195 &   0.118 &  0.02829 &  0.2853  &    -0.1153 &  0.4987& -0.26 &1.06737&0.996 &0.12334& 0.969\\
   0.59  &   0.436 &   1.029 &   0.0108 &   0.145 &  0.02691 &  0.2554  &    -0.0949 &  0.4985& -0.02 &1.03198&0.997 &0.01472&0.734\\
   0.59  &   0.410 &   1.008 &  -0.0417 &   0.159 &  0.02592 &  0.2404  &    -0.0853 &  0.4991& 0.09 & 1.00774&1.000&-0.04012&1.039\\
   0.59  &   0.359 &   0.943 &  -0.1534 &   0.188 &  0.02342 &  0.2104  &    -0.0673 &  0.4989& 0.37 &0.94638 & 0.996 &-0.14830&1.034\\
   %%0.59  &   0.282 &   0.800 &  -0.3307 &   0.235 &  0.01849 &  0.1648  &    -0.0433 &  0.4950& 0.94 &0.80519& 0.994 &-0.32384&1.021\\
   0.59  &   0.231 &   0.584 &  -0.5426 &   0.267 &  0.01475 &  0.1351  &    -0.0300 &  0.4910& 2.1 & 0.60698& 0.962&-0.51563&1.013\\
   \tableline

   0.41  &   0.385 &   0.971 &  0.1369  &   0.132 &  0.01220 &  0.1563  &    -0.0407 &  0.4978& -0.40 & 0.97108& 1.000&0.14159&0.967\\
   0.41  &   0.333 &   0.930 &  0.0016  &   0.165 &  0.01112 &  0.1347  &    -0.0313 &  0.4998& -0.005 &0.92963& 1.000 &0.00331&0.483\\
   0.41  &   0.308 &   0.900 & -0.0685  &   0.183 &  0.01061 &  0.1254  &    -0.0275 &  0.4991& 0.22 &0.89982&1.000 &-0.06224&1.101\\
   0.41  &   0.256 &   0.808 & -0.2074  &   0.217 &  0.00896 &  0.1046  &    -0.0199 &  0.4937& 0.73 &0.81275&0.994 &-0.20138&1.030\\
   %%%%0.41  &   0.205 &   0.668 & -0.3715  &   0.252 &  0.00695 &  0.0838  &    -0.0132 &  0.4868 & &\\
   \tableline
   0.28  &   0.256 &   0.809 &  0.0246  &   0.183 &  0.00472 &  0.0717  &    -0.0103 &  0.4982&-0.12 &0.80944&0.999 &0.03668&0.671\\
   0.28  &   0.231 &   0.780 & -0.0572  &   0.204 &  0.00431 &  0.0646  &    -0.0086 &  0.4988&0.28 & 0.77651&1.004&-0.04714&1.213\\
   0.28  &   0.205 &   0.726 & -0.1505  &   0.224 &  0.00382 &  0.0575  &    -0.0069 &  0.4967&0.80 & 0.72853&0.996&-0.13511&1.114\\
%%  \tableline
%%\end{tabular}
%%\end{center}
%%\end{table}
\enddata
\end{deluxetable}
%\end{landscape}

\end{rotate}

\begin{deluxetable}{ccccccccccc}
\tablecolumns{11}
\tablewidth{0pt}
\tablecaption{Direct Configurations with $n=0.5$\label{n05}}
\tablehead{
\colhead{$b/a$} & \colhead{$c/a$} & \colhead{$\omega$} &
\colhead{$\lambda$} & \colhead{$T/|W|$} & \colhead{$J_{\mathrm{tot}}$} &
\colhead{$M_{\mathrm{tot}}$} & \colhead{$\rho_{\mathrm{mean}}$} &
\colhead{$W$} & \colhead{$(S+T)/|W|$} & \colhead{$f$} 
} \startdata
                                                                                                                                     
%%\begin{table}[]
%%\begin{center}
%%\caption{Direct Configurations for Compressible counterparts of Riemann S-type Ellipsoids for $n=0.5$\label{n05}}
%%\begin{tabular}{|c|c|c|c|c|c|c|c|c|c|c|}
%%  \tableline
%%   $b/a$ & $c/a$ & $\omega$ & $\lambda$ & $T/|W|$ & $J_{\mathrm{tot}}$ &
%%   $M_{\mathrm{tot}}$ & $\rho_{\mathrm{mean}}$ & $W$ & $(S+T)/|W|$ & $f$ \\ 
%%   \tableline

   0.90  &  0.795  & 0.970 &   0.3660 &   0.045 &  0.0258 & 0.3812 &  0.5398  &    -0.1751  &  0.5000& -0.76\\
   0.90  &  0.744  & 0.975 &   0.2831 &   0.061 &  0.0273 & 0.3541 &  0.5371  &    -0.1544  &  0.5000& -0.58\\
   0.90  &  0.692  & 0.970 &   0.2050 &   0.079 &  0.0276 & 0.3266 &  0.5344  &    -0.1344  &  0.5001&-0.43\\
   0.90  &  0.641  & 0.958 &   0.1312 &   0.097 &  0.0227 & 0.2987 &  0.5309  &    -0.1150  &  0.5001& -0.28\\
   0.90  &  0.590  & 0.940 &   0.0617 &   0.117 &  0.0256 & 0.2695 &  0.5263  &    -0.0964  &  0.5001& -0.13\\
   0.90  &  0.564  & 0.929 &   0.0287 &   0.127 &  0.0246 & 0.2546 &  0.5240  &    -0.0873  &  0.5001& -0.06\\
   *0.90  &  0.538  & 0.920 &   0.0001 &   0.137 &  0.0233 & 0.2391 &  0.5198  &    -0.0783  &  0.5001& -0.0002\\
   0.90  &  0.513  & 0.908 &  -0.0281 &   0.147 &  0.0218 & 0.2234 &  0.5169  &    -0.0695  &  0.5001& 0.06\\
   0.90  &  0.487  & 0.898 &  -0.0513 &   0.157 &  0.0201 & 0.2066 &  0.5115  &    -0.0607  &  0.5001& 0.12\\
   \tableline
   0.74  &  0.692  & 0.958 &   0.3232 &   0.055 &  0.0166 & 0.2719 &  0.5380  &    -0.0991  &  0.4997& -0.71\\
   0.74  &  0.641  & 0.959 &   0.2343 &   0.074 &  0.0171 & 0.2489 &  0.5345  &    -0.0852  &  0.4999& -0.51\\
   0.74  &  0.590  & 0.949 &   0.1514 &   0.094 &  0.0167 & 0.2252 &  0.5314  &    -0.0717  &  0.5001& -0.33\\
   0.74  &  0.538  & 0.932 &   0.0750 &   0.114 &  0.0156 & 0.2005 &  0.5263  &    -0.0587  &  0.5001&-0.17\\
   *0.74  &  0.487  & 0.911 &   0.0081 &   0.135 &  0.0137 & 0.1737 &  0.5192  &    -0.0458  &  0.5002& -0.02\\
   0.74  &  0.462  & 0.904 &  -0.0171 &   0.144 &  0.0123 & 0.1588 &  0.5156  &    -0.0393  &  0.5002& 0.04\\
   \tableline
   0.59  &  0.718  & 0.792 &   0.6631 &   0.012 &  0.0037 & 0.2241 &  0.5418  &    -0.0713  &  0.4943& -1.9\\
   0.59  &  0.538  & 0.922 &   0.2021 &   0.085 &  0.0090 & 0.1587 &  0.5304  &    -0.0397  &  0.4997& -0.50\\
   0.59  &  0.487  & 0.911 &   0.1137 &   0.105 &  0.0081 & 0.1373 &  0.5256  &    -0.0313  &  0.5001& -0.29\\
   0.59  &  0.462  & 0.905 &   0.0754 &   0.114 &  0.0073 & 0.1250 &  0.5230  &    -0.0265  &  0.5002& -0.19\\
   0.59  &  0.436  & 0.871 &  -0.0337 &   0.153 &  0.0105 & 0.1382 &  0.5453  &    -0.0031  &  0.5088& 0.09\\
%%  \tableline
%%\end{tabular}
%%\end{center}
%%\end{table}

\enddata
\end{deluxetable}

\begin{deluxetable}{ccccccccccc}
\tablecolumns{11}
\tablewidth{0pt}
\tablecaption{Direct Configurations with $n=1.0$\label{n10}}
\tablehead{
\colhead{$b/a$} & \colhead{$c/a$} & \colhead{$\omega$} &
\colhead{$\lambda$} & \colhead{$T/|W|$} & \colhead{$J_{\mathrm{tot}}$} &
\colhead{$M_{\mathrm{tot}}$} & \colhead{$\rho_{\mathrm{mean}}$} &
\colhead{$W$} & \colhead{$(S+T)/|W|$} & \colhead{$f$}
} \startdata

%%\begin{table}[]
%%\begin{center}
%%\caption{Direct Configurations for Compressible Counterparts of Riemann S-type Ellipsoids for $n=1.0$\label{n10}}
%%\begin{tabular}{|c|c|c|c|c|c|c|c|c|c|c|}
%%  \tableline
%%   $b/a$ & $c/a$ & $\omega$ & $\lambda$ & $T/|W|$ & $J_{\mathrm{tot}}$ &
%%   $M_{\mathrm{tot}}$ & $\rho_{\mathrm{mean}}$ & $W$ & $(S+T)/|W|$ & $f$\\
%%   \tableline
   0.90  &  0.795  & 0.803  & 0.3029 &  0.039 & 0.00893 & 0.2053 & 0.2921  &    -0.0577 &    0.5001 &-0.76\\
   0.90  &  0.744  & 0.804  & 0.2329 &  0.053 & 0.00910 & 0.1870 & 0.2867  &    -0.0492 &    0.5001 &-0.58\\
   0.90  &  0.692  & 0.796  & 0.1684 &  0.067 & 0.00872 & 0.1679 & 0.2799  &    -0.0409 &    0.5002 &-0.43\\
   0.90  &  0.641  & 0.782  & 0.1090 &  0.081 & 0.00788 & 0.1476 & 0.2711  &    -0.0328 &    0.5002 &-0.28\\
   0.90  &  0.590  & 0.760  & 0.0555 &  0.093 & 0.00657 & 0.1255 & 0.2611  &    -0.0249 &    0.5002 &-0.14\\
   0.90  &  0.564  & 0.747  & 0.0314 &  0.097 & 0.00572 & 0.1134 & 0.2563  &    -0.0210 &    0.5002 &-0.08\\
   \tableline
   0.74  &  0.641  & 0.784  & 0.1917 &  0.060 & 0.00513 & 0.1262 & 0.2805  &    -0.0254 &    0.5003 &-0.51\\
   0.74  &  0.590  & 0.769  & 0.1241 &  0.074 & 0.00450 & 0.1087 & 0.2731  &    -0.0197 &    0.5003 &-0.34\\
   0.74  &  0.538  & 0.743  & 0.0626 &  0.084 & 0.00349 & 0.0893 & 0.2680  &    -0.0143 &    0.5003 &-0.18\\
   0.74  &  0.512  & 0.722  & 0.0322 &  0.086 & 0.00289 & 0.0786 & 0.2676  &    -0.0115 &    0.5003 &-0.09\\
   \tableline
   0.59  &  0.590  & 0.748  & 0.244  & 0.046  & 0.00237 & 0.0870 & 0.2836  &    -0.0136 &    0.5007 &-0.75\\
   0.59  &  0.564  & 0.743  & 0.2011 & 0.052  & 0.00220 & 0.0797 & 0.2817  &    -0.0118 &    0.5009 &-0.62\\
%%  \tableline
%%\end{tabular}
%%\end{center}
%%\end{table}

\enddata
\end{deluxetable}

\begin{deluxetable}{ccccccccccc}
\tablecolumns{11}
\tablewidth{0pt}
\tablecaption{Direct Configurations with $n=1.5$\label{n15}}
\tablehead{
\colhead{$b/a$} & \colhead{$c/a$} & \colhead{$\omega$} &
\colhead{$\lambda$} & \colhead{$T/|W|$} & \colhead{$J_{\mathrm{tot}}$} &
\colhead{$M_{\mathrm{tot}}$} & \colhead{$\rho_{\mathrm{mean}}$} &
\colhead{$W$} & \colhead{$(S+T)/|W|$} & \colhead{$f$}
} \startdata

   0.90  &   0.795 &  0.631  &  0.2351 &   0.030 &  0.00273 & 0.1068 &  0.1531  &    -0.0181  &   0.5002&-0.75\\
   0.90  &   0.744 &  0.622  &  0.1752 &   0.040 &  0.00260 & 0.0945 &  0.1467  &    -0.0147  &   0.5003&-0.57\\
   0.90  &   0.692 &  0.604  &  0.1187 &   0.048 &  0.00229 & 0.0817 &  0.1397  &    -0.0115  &   0.5003&-0.40\\
   0.90  &   0.641 &  0.574  &  0.0640 &   0.055 &  0.00183 & 0.0683 &  0.1324  &    -0.0085  &   0.5003&-0.22\\
   0.90  &   0.615 &  0.554  &  0.0365 &   0.057 &  0.00157 & 0.0614 &  0.1292  &    -0.00711 &   0.5003&-0.13\\
   \tableline
   0.74  &   0.692 &   0.609 &  0.2024 &   0.033 &   0.00148 & 0.0717 & 0.1491  &    -0.00931 &   0.5006&-0.70\\
   0.74  &   0.641 &   0.592 &  0.1378 &   0.041 &   0.00130 & 0.0612 & 0.1430  &    -0.00713 &   0.5005&-0.49\\
   \tableline
   0.64  &   0.670 &   0.587 &  0.2503 &   0.023 &   0.00086 & 0.0586 & 0.1524  &    -0.00666 &  0.5014&-0.94\\
   0.64  &   0.640 &   0.584 &  0.2116 &   0.027 &   0.00084 & 0.0542 & 0.1499  &    -0.00584 &  0.5013&-0.80\\
\enddata
\end{deluxetable}

\begin{deluxetable}{ccccccccccc}
\tablecolumns{11}
\tablewidth{0pt}
\tablecaption{Adjoint Configurations of Riemann S-type Ellipsoids ($n=0$)\label{n0a}}
\tablehead{
\colhead{$b/a$} & \colhead{$c/a$} & \colhead{$\omega$} &
\colhead{$\lambda$} & \colhead{$T/|W|$} & \colhead{$J_{\mathrm{tot}}$} &
\colhead{$M_{\mathrm{tot}}$} & \colhead{$\rho_{\mathrm{mean}}$} &
\colhead{$W$} & \colhead{$(S+T)/|W|$} & \colhead{$f$}
} \startdata

%%\begin{table}[]
%%\begin{center}
%%\caption{Adjoint Configurations of Riemann S-type Ellipsoids ($n=0$)\label{n0a}}
%%\begin{tabular}{|c|c|c|c|c|c|c|c|c|c|c|c|}
%%  \tableline
%%
%%   $b/a$ & $c/a$ & $\omega$ & $\lambda$ & $T/|W|$ & $J_{\mathrm{tot}}$ &
%%   $M_{\mathrm{tot}}$ & $\rho_{\mathrm{mean}}$ & $W$ & $(S+T)/|W|$ &$f$\\
%%  \tableline
   0.90  &  0.923 & -0.755 & -1.030 &0.0064  & 0.0307 & 0.8232 & 0.99999  &-0.6989   &   0.4987    &-2.744\\
   0.90  &  0.641 &  -0.139& -1.119 & 0.100  & 0.0770 & 0.572  & 0.99999 & -0.3758  &   0.4980    & -16.1\\
   0.90  &  0.590 & -0.0569& -1.099 & 0.121  & 0.0754 & 0.5261 & 0.99999  & -0.3249  &   0.4979    & -38.8\\
   0.90  &  0.564 & -0.0211& -1.090 & 0.132  & 0.0741 & 0.5033 & 0.99999 & -0.3007  &   0.4981    & -104.2\\
   0.90  &  0.538 &  0.0250& -1.069 & 0.143  & 0.0723 & 0.4801 & 0.99999 & -0.2768  &   0.4982    & 86.0\\
   0.90  &  0.487 &  0.121 & -1.015 & 0.166  & 0.0678 & 0.4342 & 0.99999  & -0.2314  &   0.4983    & 16.9\\
   0.90  &  0.436 &  0.219 & -0.945 & 0.191  & 0.0623 & 0.3886 & 0.99999  & -0.1900  &   0.4979    & 8.689\\
   %%0.90  &  0.385 &  0.274 & -0.909 & 0.217  & 0.0559 & 0.3431 & 0.99999  & -0.1518  &   0.4980    & 6.682\\
   0.90  &  0.333 &  0.389 & -0.797 & 0.246  & 0.0485 & 0.2970 & 0.99999  & -0.1167  &   0.4978    & 4.128\\
  \tableline
   0.74  &  0.744 & -0.495 & -1.117 & 0.035  & 0.0376 & 0.5496 & 0.99999 & -0.3540   &  0.4943  &   -4.7\\
   0.74  &  0.590 & -0.171 & -1.116 & 0.093  & 0.0465 & 0.4357 & 0.99999 & -0.2377   &  0.4933  &   -13.6\\
   0.74  &  0.513 & -0.030 & -1.078 &  0.127 & 0.0452 & 0.3791 & 0.99999  & -0.1866   &  0.4932  &   -74.3\\
   0.74  &  0.487 &  0.021 & -1.054 &  0.139 & 0.0440 & 0.3595 & 0.99999  & -0.1699   &  0.4928  &   106.8\\
   0.74  &  0.436 &  0.118 & -1.002 &  0.165 & 0.0414 & 0.3223 & 0.99999  & -0.1400   &  0.4929  &   17.7\\
   0.74  &  0.385 &  0.210 & -0.944 &  0.193 & 0.0376 & 0.2841 & 0.99999  & -0.1117   &  0.4934  &   9.4\\
   0.74  &  0.333 &  0.323 & -0.847 &  0.223 & 0.0332 & 0.2463 & 0.99999  & -0.0863   &  0.4940  &   5.5\\
   0.74  &  0.282 &  0.453 & -0.714 &  0.257 & 0.0281 & 0.2081 & 0.99999  & -0.0634   &  0.4944  &   3.4\\
  \tableline
   0.59  &  0.615 & -0.417 & -1.085 &  0.045 & 0.0199 & 0.3608 & 0.99999  & -0.1729   &  0.4834  &   -5.5\\
   0.59  &  0.487 & -0.119 & -1.063 &  0.100 & 0.0239 & 0.2853 & 0.99999  & -0.1152   &  0.4807  &   -20.4\\
   0.59  &  0.436 & -0.011 & -1.029 &  0.126 & 0.0235 & 0.2553 & 0.99999  & -0.0949   &  0.4801  &  -217.5\\
   0.59  &  0.410 &  0.042 & -1.008 &  0.140 & 0.0229 & 0.2404 & 0.99999  & -0.0853   &  0.4807  &  55.2\\
   0.59  &  0.359 &  0.153 & -0.943 &  0.170 & 0.0213 & 0.2104 & 0.99999  & -0.0672   &  0.4811  &  14.1\\
   0.59  &  0.282 &  0.271 & -0.856 &  0.204 & 0.0190 & 0.1803 & 0.99999  & -0.0509   &  0.4822  &  7.2\\
   0.59  &  0.231 &  0.402 & -0.729 &  0.242 & 0.0161 & 0.1502 & 0.99999  & -0.0365   &  0.4843  &  4.1\\
%%  \tableline
%%\end{tabular}
%%\end{center}
%%\end{table}

\enddata
\end{deluxetable}

\begin{deluxetable}{ccccccccccc}
\tablecolumns{11}
\tablewidth{0pt}
\tablecaption{Adjoint Configurations with $n=0.5$ \label{n05a}}
\tablehead{
\colhead{$b/a$} & \colhead{$c/a$} & \colhead{$\omega$} &
\colhead{$\lambda$} & \colhead{$T/|W|$} & \colhead{$J_{\mathrm{tot}}$} &
\colhead{$M_{\mathrm{tot}}$} & \colhead{$\rho_{\mathrm{mean}}$} &
\colhead{$W$} & \colhead{$(S+T)/|W|$} & \colhead{$f$}
} \startdata

%%\begin{table}[]
%%\begin{center}
%%\caption{Adjoint Configurations of Riemann S-type Ellipsoids ($n=0.5$)\label{n05a}}
%%\begin{tabular}{|c|c|c|c|c|c|c|c|c|c|c|c|}
%%  \tableline
%%   $b/a$ & $c/a$ & $\omega$ & $\lambda$ & $T/|W|$ & $J_{\mathrm{tot}}$ &
%%   $M_{\mathrm{tot}}$ & $\rho_{\mathrm{mean}}$ & $W$ & $(S+T)/|W|$ &$f$\\
%%  \tableline

   0.90  &  0.923 &-0.642  &-0.8766  &0.0063 &0.0116  &0.4476  &0.5442   &  -0.2297 &   0.4997  &   -2.7\\
   0.90  &  0.590 &-0.062  &-0.9401  &0.116  &0.0255  &0.2695  &0.5263   &  -0.0964 &   0.4997  &  -30.7\\
   0.90  &  0.538 &-1.47e-4&-0.9195  &0.137  &0.0232  &0.2391  &0.5197   &  -0.0782 &   0.4999  &  -12606.1\\
   0.90  &  0.513 & 0.028  &-0.9080  &0.147  &0.0218  &0.2234  &0.5169   &  -0.0695 &   0.5000  &  65.2\\
  \tableline
   0.74  &  0.795 & -0.537 & -0.9101 & 0.018 & 0.0102  &0.3167 & 0.5421  &  -0.1283 &    0.4966 &   -3.5\\
   0.74  &  0.692 & -0.323 & -0.9584 & 0.052 & 0.0154  &0.2719 & 0.5379  &  -0.0991 &    0.4963 &   -6.2\\
   0.74  &  0.538 & -0.075 & -0.9317 & 0.112 & 0.0150  &0.2005 & 0.5263  &  -0.0587 &    0.4975 &   -25.9\\
   0.74  &  0.487 & -8.2e-3& -0.9114 & 0.134 & 0.0133  &0.1737 & 0.5192  &  -0.0458 &    0.4992 &   -233.1\\
   0.74  &  0.462 & 0.0171 & -0.9044 & 0.145 & 0.0121  &0.1587 & 0.5157  &  -0.0393 &    0.5009 &   110.2\\
  \tableline
   0.59  &  0.615 & -0.355 & -0.9180 & 0.044 & 6.93e-3 &0.1877 & 0.5368  &   -0.0529&    0.4894 &   -5.9\\
   0.59  &  0.538 & -0.202 & -0.9222 & 0.076 & 7.66e-3 &0.1587 & 0.5304  &   -0.0397&    0.4910 &   -10.4\\
   0.59  &  0.487 & -0.114 & -0.9111 & 0.100 & 7.21e-3 &0.1373 & 0.5256  &   -0.0310&    0.4948 &   -18.3\\
   0.59  &  0.462 & -0.075 & -0.9050 & 0.113 & 6.65e-3 &0.1250 & 0.5230  &   -0.0265&    0.4992 &   -27.4\\
%%  \tableline
%%\end{tabular}
%%\end{center}
%%\end{table}

\enddata
\end{deluxetable}

\begin{deluxetable}{ccccccccccc}
\tablecolumns{11}
\tablewidth{0pt}
\tablecaption{Adjoint Configurations with $n=1.0$\label{n10a}}
\tablehead{
\colhead{$b/a$} & \colhead{$c/a$} & \colhead{$\omega$} &
\colhead{$\lambda$} & \colhead{$T/|W|$} & \colhead{$J_{\mathrm{tot}}$} &
\colhead{$M_{\mathrm{tot}}$} & \colhead{$\rho_{\mathrm{mean}}$} &
\colhead{$W$} & \colhead{$(S+T)/|W|$} & \colhead{$f$}
} \startdata

   0.90  &  0.795  & -0.303  & -0.8030 &  0.039 & 0.00887 & 0.2053 & 0.2921  &    -0.0577 &    0.5000 &-5.33\\
   0.90  &  0.692  & -0.168  & -0.7963 &  0.067 & 0.00869 & 0.1679 & 0.2799  &    -0.0409 &    0.5003 &-9.51\\
   0.90  &  0.590  & -0.056  & -0.7605 &  0.094 & 0.00657 & 0.1255 & 0.2611  &    -0.0249 &    0.5009 &-27.59\\
   0.90  &  0.564  & -0.031  & -0.7469 &  0.099 & 0.00573 & 0.1134 & 0.2563  &    -0.0210 &    0.5014 &-47.78\\
   \tableline
   0.74  &  0.641  & -0.192  & -0.7842 &  0.061 & 0.00499 & 0.1262 & 0.2805  &    -0.0254 &    0.5013 &-8.54\\
   0.74  &  0.538  & -0.062  & -0.7429 &  0.091 & 0.00354 & 0.0893 & 0.2680  &    -0.0142 &    0.5071 &-24.80\\
   0.74  &  0.512  & -0.032 & -0.7220 &  0.096 & 0.00295 & 0.0786 & 0.2676  &    -0.0115 &    0.5097 &-46.74\\
   \tableline
   0.59  &  0.590  & -0.244  & -0.7480  & 0.050  & 0.00217 & 0.0870 & 0.2836  &    -0.0136 &    0.5049 &-7.01\\
   0.59  &  0.564  & -0.201& -0.7433 & 0.059  & 0.00210 & 0.0797 & 0.2817  &    -0.0118 &    0.5083 &-8.47\\

\enddata
\end{deluxetable}

%\end{spacing}

%% The following command ends your manuscript. LaTeX will ignore any text
%% that appears after it.


\begin{thebibliography}{}
%%\bibitem[Andersson(2003)]{A03} Andersson, N. 2003, Class. Quant. Grav., 20, 105
\bibitem[Binney \& Tremaine(1987)]{BT87}Binney, J., \& Tremaine, S., 1987, 
Galactic Dynamics, New York: Academic Press
\bibitem[Cazes \& Tohline(2000)]{CT00}Cazes, J. E., \& Tohline, J. E. 2000, \apj, 532, 1051
\bibitem[Chandrasekhar(1969)]{Ch69}Chandrasekhar, S. 1969,
Equilibrium Figures of Equilibrium,  New Haven, CT:  Yale Univ.
Press

%%\bibitem[Chandrasekhar(1970)]{Ch70} Chandrasekhar, S. 1970, \apj,
%%161, 561

\bibitem[Cohl \& Tohline(1999)]{CT99}Cohl, H. S., \& Tohline, J. E. 1999, \apj, 527, 86
%%\bibitem[Durisen et al.(1986)]{DGTB86} Durisen, R. H., Gingold, R. A., Tohline, J. E., \& Boss, A. P.
%%1986, \apj, 305, 281

\bibitem[Friedman \& Schutz(1978)]{FS78} Friedman, J., \& Schutz,
B. F. 1978, \apj, 222, 281
%%\bibitem[Gonz\'alez(2004)]{G04}Gonz\'alez, G. 2004, ``Status of LIGO
%%Data Analysis,'' in Proceedings of the $8^{\mathrm{th}}$
%%Gravitational Wave Data Analysis Workshop, Class. Quant. Grav.,
%%submitted
\bibitem[Hachisu(1986a)]{H86A} Hachisu, I. 1986, \apjs, 61, 479
\bibitem[Hachisu(1986b)]{H86B} Hachisu, I. 1986, \apjs, 62, 461
\bibitem[Hachisu \& Eriguchi(1982)]{HE82} Hachisu, I., Eriguchi, Y., 1982, Prog. Theor.Phys., 68, 206
\bibitem[Hachisu \& Eriguchi(1984)]{HE84} Hachisu, I., Eriguchi, Y., 1984, \pasj, 36, 239
%%\bibitem[Hawley, Balbus \& Winters(1999)]{HBW99} Hawley, J. F.,
%%Balbus, S. A., \& Winters, W. F. 1999, \apj, 518, 394
%%\bibitem[Imamura, Friedman, \& Durisen(1985)]{IFD85} Imamura, J.
%%N., Friedman, J. L., \& Durisen, R. H. 1985, \apj, 294, 474
%%\bibitem[Ipser \& Lindblom(1990)]{IL90} Ipser, J. R., \&
%%Lindblom, L. 1990, \apj, 355, 226
%%\bibitem[Ipser \& Lindblom(1991)]{IL91} Ipser, J. R., \&
%%Lindblom, L. 1991, \apj, 373, 213

\bibitem[James(1964)]{J64} James, R. A., 1964, \apj, 140, 552

\bibitem[Lai, Rasio \& Shapiro(1993)]{LRS93} Lai, D., Rasio, F. A.,  \& Shapiro, S. L.
1993, \apjs, 88, 205
\bibitem[Lai \& Shapiro(1995)]{LS95} Lai, D., \& Shapiro, S. L.
1995, \apj, 442, 259


\bibitem[Lebovitz \& Lifschitz(1996)]{LL96}Lebovitz, N. R., \& Lifschitz, A. 1996, \apj, 458, 699
%%"New Global Instabilities of the Riemann Ellipsoids"

\bibitem[Lebovitz \& Saldanha(1999)]{LS99}Lebovitz, N. R., \& Saldanha, K. I. 1999, Physics of Fluids,
11, 3374
%%"On the weakly nonlinear development of the elliptic instability"

\bibitem[Lifschitz \& Lebovitz(1993)]{LL93}Lifschitz, A., \& Lebovitz, N. 1993, \apj, 408, 603
%%"Local hydrodynamic instability of rotating stars"

%%\bibitem[Managan(1985)]{M85} Managan, R. A. 1985, \apj, 294, 463
\bibitem[Motl, Tohline, \& Frank(2002)]{MTF02} Motl, P. M.,
Tohline, J. E., \& Frank, J. 2002, \apjs, 138, 121


%%\bibitem[New, Centrella \& Tohline(2000)]{NCT00}New, K. C. B., Centrella, J. M., \& Tohline, J. E.
%%2000, \prd, 62, 064019
%%\bibitem[Stergioulas(2003)]{S03}Stergioulas, N. 2003, Living Rev. Relativity, 6, 3
%%[Online article]: cited on 26 April 2004
%%http://www.livingreviews.org/lrr-2003-3/

\bibitem[Ostriker \& Mark(1968)]{OM68} Ostriker, J.P.,
Mark, J. W.-K., 1968, \apj, 151, 1075

\bibitem[Ou, Tohline, \& Lindblom(2004)]{OTL04} Ou, S.,
Tohline, J. E., \& Lindblom, L. 2004, \apj, 617, 490

\bibitem[Shibata \& Karino(2004)]{SK04}Shibata, M., \& Karino, S., 2004, \prd, 70, 084022

\bibitem[Shibata, Karino, \& Eriguchi(2002)]{SKE02}Shibata, M., Karino, S., \& Eriguchi, Y. 2002, \mnras, 334,L27

%%\bibitem[Tassoul(1978)]{T78}Tassoul, J.-L. 1978, Theory of Rotating Stars,
%%Princeton: Princeton University Press
\bibitem[Tohline, Durisen, \& McCollough(1985)]{TDM85}Tohline, J. E., Durisen, R. H., \& McCollough, M.
1985, \apj, 298, 220


\bibitem[Uryu \& Eriguchi(1996)]{UE96}Uryu, K., \& Eriguchi, Y., 1996, \mnras, 282, 653
\bibitem[Uryu \& Eriguchi(1998)]{UE98}Uryu, K., \& Eriguchi, Y., 1998, \apjs, 118, 563

\bibitem[Williams \& Tohline(1988)]{WT88}Williams, H. A., \& Tohline, J. E. 1988, \apj, 334, 449
\end{thebibliography}
\end{document}